\documentclass[amsmath,trackchanges,twocolumn]{aastex702}

\newcommand{\lessthan}{\textless \space}
\newcommand{\greaterthan}{\textgreater \space}

\newcommand{\kms}{km~s$^{-1}$}
\newcommand{\kmspace}{km~s$^{-1}$\space}

\newcommand{\dexkpc}{dex~kpc$^{-1}$}
\newcommand{\degrees}{$^{\circ}$}
\newcommand{\degreespace}{$^{\circ}$\space}

\defcitealias{Gibson2023}{G23}
\defcitealias{Gibson2025}{G25}

\usepackage{booktabs}

\begin{document}

\title{Probing the Assembly of Triangulum: Precision Metallicities and $\alpha$ Abundances of M33's Star Clusters from APOGEE Integrated-Light Spectroscopy}

\author[orcid=0000-0001-8203-6004,sname='Gibson']{Benjamin J. Gibson}
\affiliation{Space Telescope Science Institute, 3700 San Martin Drive, Baltimore, MD 21218, USA}
\affiliation{Department of Physics and Astronomy, University of Utah, 115 S 1400 E, Salt Lake City, UT 84112}
\email[show]{bengibson.astro@gmail.com}  

\author[orcid=0000-0001-5261-4336,sname='Anguiano']{Borja Anguiano} 
\altaffiliation{Ramon y Cajal Fellow}
\affiliation{Centro de Estudios de Física del Cosmos de Aragón (CEFCA), Plaza San Juan 1, 44001, Teruel, Spain}
\affiliation{Department of Astronomy, University of Virginia, Charlottesville, VA. 22904-4325, USA}
\email{astrobaj@gmail.com}

\author[orcid=0000-0001-5388-0994,sname='Hasselquist']{Sten Hasselquist} 
\affiliation{Space Telescope Science Institute, 3700 San Martin Drive, Baltimore, MD 21218, USA}
\email{shasselquist@stsci.edu}

\author[orcid=0000-0001-6761-9359,sname='Zasowski']{Gail Zasowski} 
\affiliation{Department of Physics and Astronomy, University of Utah, 115 S 1400 E, Salt Lake City, UT 84112}
\email{u0948422@gcloud.utah.edu}

\author[orcid=0000-0003-0248-5470,sname='Seth']{Anil Seth} 
\affiliation{Department of Physics and Astronomy, University of Utah, 115 S 1400 E, Salt Lake City, UT 84112}
\email{aseth@astro.utah.edu}

\author[orcid=0000-0002-1691-8217,sname='Beaton']{Rachael L. Beaton} 
\affiliation{Space Telescope Science Institute, 3700 San Martin Drive, Baltimore, MD 21218, USA}
\email{rbeaton@stsci.edu}

\author[orcid=0000-0003-2969-2445,sname='Hayes']{Christian R. Hayes} 
\affiliation{Space Telescope Science Institute, 3700 San Martin Drive, Baltimore, MD 21218, USA}
\email{chayes@stsci.edu}

\author[orcid=0000-0003-3081-9319,sname='Lewis']{Geraint F. Lewis} 
\affiliation{Sydney Institute for Astronomy, School of Physics, A28, The University of Sydney, NSW 2006, Australia}
\email{geraint.lewis@sydney.edu.au}

\author[orcid=0000-0003-2025-3147,sname='Majewski']{Steven R. Majewski} 
\affiliation{Department of Astronomy, University of Virginia, Charlottesville, VA. 22904-4325, USA}
\email{srm4n@virginia.edu}

\author[orcid=0009-0005-6518-3914,sname='Pagnini']{Giulia Pagnini} 
\affiliation{Université de Strasbourg, CNRS, Observatoire Astronomique de Strasbourg, UMR 7550, F-67000 Strasbourg, France}
\affiliation{Centre national d’études spatiales (CNES), 2, Place Maurice Quentin, 75039, Paris, France}
\email{pagnini@astro.unistra.fr}

\begin{abstract}

The star cluster system of M33 provides an excellent opportunity to investigate whether galaxy formation physics is dependent on a galaxy's mass. Older globular clusters (GCs) trace hierarchical assembly and chemical enrichment, while younger clusters trace the recent star formation. By studying the M33 star cluster system as a whole, we can probe the formation physics of galaxies at a lower mass than the Milky Way, and look for signals of accretion. To that end, we present analysis of 37 M33 star clusters using high-resolution ($R \sim 22,500$), near-infrared (1.5-1.7 $\mu$m), integrated light spectra from the APOGEE survey. We use full-spectrum fitting to obtain precision measurements of the clusters' kinematics, metallicities, and $\alpha$ abundances. We identify three populations of clusters from our measurements: young stellar clusters ($\sim$100 Myr) with broad Brackett absorption features; intermediate-age clusters (1-2 Gyr) that are metal-rich ([M/H] = -0.5 to +0.3) with roughly solar $\alpha$ abundance; and old ($>$5 Gyr), kinematically-hot and metal-poor GCs. These old GCs exhibit a broad range of kinematics (RV $\sim$ -250 to -8 \kms), metallicities ([M/H] = -1.8 to -0.6) and $\alpha$ abundances ($\rm [\alpha/M]$ = +0.05 to +0.3), and we identify and discuss three GCs that show strong chemodynamical evidence of accretion into M33's halo. M33's GCs at [M/H] $\sim$ -1.2 have a spread in [$\alpha$/M] of nearly 0.2 dex. This pattern is strikingly similar to the abundance distribution of Milky Way GCs, and indicates either that M33 and the Milky Way formed clusters similarly at early times, or that M33 hosts a significant population of accreted clusters in its halo. 

\end{abstract}


\section{Introduction}\label{sec:intro}

Hierarchical galaxy formation in a $\Lambda$CDM universe predicts that galaxies assemble through a combination of \emph{in-situ} star formation and the accretion of smaller satellite systems \citep[e.g.,][]{WF1991,Mo1998}. Consequently, the stellar halos of massive galaxies are expected to comprise populations with multiple origins, including stars formed \textit{in-situ} then dynamically heated into the halo, as well as stars accreted during galaxy mergers. Observations of nearby galaxies support this picture, revealing that mergers play a fundamental role in shaping galaxy morphology and stellar populations \citep[e.g.,][]{Martinez2019}. 

In the Milky Way (MW), numerous stellar streams and kinematically distinct halo populations have been identified as the debris of past accretion events \citep{Ibata1995,Helmi1999,Majewski2003,Belokurov2006,Myeong2019}. Many globular clusters (GCs) have been confidently associated with specific progenitor galaxies, such as the Sagittarius dwarf spheroidal and the Gaia-Sausage/Enceladus merger \citep[e.g.,][and references therein]{Massari2019}. 

GCs are powerful tracers of hierarchical   assembly, as they a) are ancient and long lived, b) preserve a fossil record of the environment in which they formed, c) can form \textit{in-situ} or be accreted, and d) are luminous and can be identified far beyond the MW \citep[e.g.,][]{Searle1978,Forbes2010,Huxor2014,Usher2024}. Additionally, younger star clusters trace the recent \textit{in-situ} star formation and dynamics of the host galaxy \citep[e.g.,][]{Adamo2020}. This makes the study of a galaxy's entire cluster population a comprehensive way to study it's past and present.

As the third most massive member of the Local Group, the Triangulum Galaxy (M33) provides an important laboratory for studying galaxy assembly beyond the Milky Way \citep[e.g.,][]{Corbelli2014}. Its evolutionary history, however, remains uncertain. Deep imaging from the Pan-Andromeda Archaeological Survey \citep[PAndAS;][]{McConnachie2009} revealed low-surface-brightness stellar substructures, streams, and GCs extending between M31 and M33, suggesting that the two galaxies may have experienced a past close encounter \citep{Cockcroft2011,McConnachie2018}. More recent proper motion measurements from the \textit{Gaia} mission and \textit{HST} instead favor a scenario in which M33 is currently undergoing its first infall toward M31 \citep{Gaia,vanderMarel2019,Patel2025}. Likewise, numerical modeling of M33's warped disk indicates that a recent close passage around M31 is unlikely to reproduce its observed morphology, instead favoring recent gas accretion as the origin of the outer disk warp \citep{Corbelli2024}. Complementing these studies, spectroscopy of individual stars have identified a kinematically hot, non-rotating stellar population that may represent an accreted halo component \citep{Gilbert2022,Cullinane2023}. Finally, \citet{Smercina2023} fit a broken power law to \textit{HST} imagery from the PHATTER survey \citep{PHATTER} and found a significant accreted halo component, making up $\sim$15.6\% of M33 total stellar mass and concentrated in the inner 2.3 kpc. Together, these results suggest that M33 has had an active interaction history, but that it remains incompletely understood.

The GC system of M33 offers an independent means of investigating this history. Early spectroscopic studies identified a cluster population spanning a wide range of ages and found that the oldest clusters exhibit halo-like kinematics with a larger velocity dispersion than the disk population \citep{CS1983,Schommer1991}. Subsequent spectroscopic surveys confirmed that the cluster system contains both ancient and young GCs, spanning ages of approximately 1--12 Gyr and metallicities ranging from [M/H] $\lesssim$-1.7 to approximately solar \citep{chandar+2002,sharina+2010,beasley+2015}. This broad age distribution distinguishes M33 from the Milky Way, whose GC system is dominated by old ($\gtrsim10$ Gyr) clusters \citep[e.g.,][]{MarinFranch2009}, and instead more closely resembles the cluster population of the Large Magellanic Cloud \citep[e.g.,][]{Geisler1997}.  Furthermore, \citet{Anguiano2025} used a radial velocity projection method to find evidence of a sub-population of M33 clusters with potentially retrograde orbits. 

In addition to its ancient globular cluster population, M33 hosts a rich population of young and intermediate-age stellar clusters distributed throughout its disk \citep[e.g.,][]{Verley2007,Moeller2022,Johnson2022,wainer2022}. These clusters are the long-lived remnants of the densest star-forming regions that survive the disruptive effects of stellar feedback and gas dispersal, and they contain a significant fraction of the stellar mass formed in nearby galaxies \citep{Kruijssen2012,Johnson2016,Adamo2020}. Optical and infrared surveys have identified hundreds of young cluster candidates in M33, including deeply embedded systems revealed through \textit{Spitzer} observations \citep{Verley2007,Sharma2011}, demonstrating that cluster formation remains an active process in the galaxy. Spectroscopic studies further indicate that these young and intermediate-age clusters represent the metal-rich end of the M33 cluster population. Their metallicities range from approximately -0.6 to +0.2 dex, depending on radius, and are broadly consistent with those of young massive stars in the present-day disk \citep{Urbaneja2005,beasley+2015}. Together, these results suggest that M33's youngest clusters provide a valuable probe of the galaxy's recent chemical enrichment and star formation history.


In this work, we present the first high-resolution near-infrared spectroscopic study of the M33 cluster system using integrated-light observations from the Apache Point Observatory Galactic Evolution Experiment \citep[APOGEE; ][]{Majewski2017}. Our sample represents the largest collection of high-resolution spectra obtained for M33 star clusters to date and complements previous optical studies by providing precise radial velocities together with homogeneous measurements of metallicities and $\alpha$ abundances. These data allow us to investigate the chemical and kinematic properties of the M33 cluster system, distinguish disk and halo populations, and identify clusters whose chemodynamical properties are consistent with an accreted origin.

This paper is organized as follows: in Section~\ref{sec:sample} we describe the observations and define our sample of bona fide clusters. In Sections~\ref{sec:dataproc}, \ref{sec:models}, and \ref{sec:fitting}, we describe our data processing, spectral models, and full-spectrum fitting routine. We present the results of our analysis in Section~\ref{sec:results} and discuss them in the context of the literature in Section~\ref{sec:discussion}, where we present a few strong candidates for accretion into M33's halo. We summarize our results and conclusions in Section~\ref{sec:conclusion}.

\section{Observations and Sample Selection} \label{sec:sample}

\begin{figure*}
\centering
\includegraphics[width=0.8\textwidth]{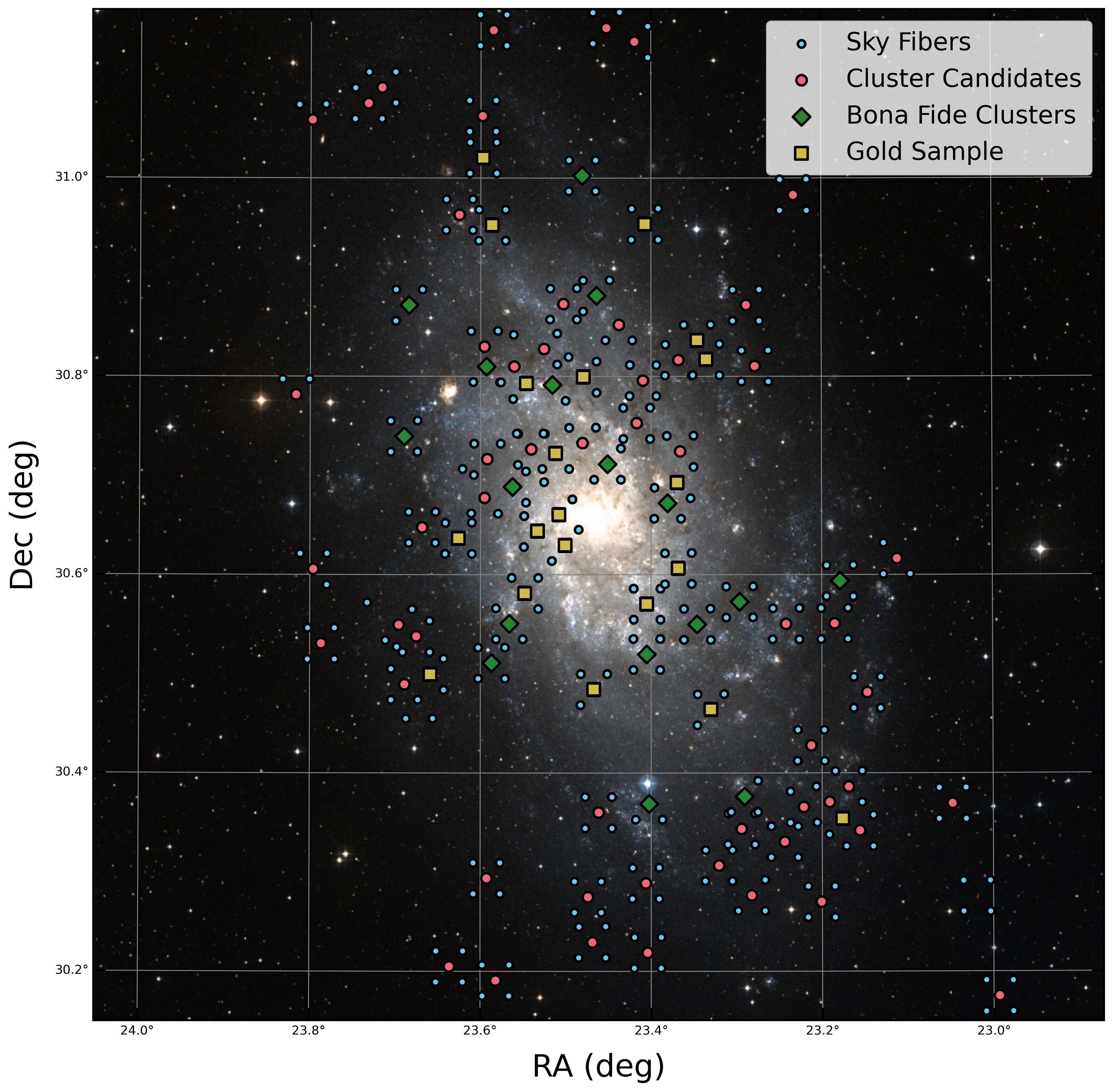}
\caption{APOGEE fiber positions in the M33 cluster sample overlaid on an optical image of M33 from the Digitized Sky Survey (DSS, Credit: Infrared Science Archive at IPAC and California Institute of Technology). The full sample of 92 cluster candidates are shown as either yellow squares, which are clusters the Gold sample (see Section~\ref{sec:bonafides}), green diamonds, which are bona fide clusters not in the Gold sample (YSCs or clusters with eS/N \textless 15), or pink circles, which are candidates that we could not robustly verify are clusters. Several of these are likely clusters, but many of these are certainly foreground MW stars. Positions of sky fibers used to subtract the sky and galaxy background are shown as small blue circles.}
\label{fig:fiber_positions}
\end{figure*}

In this section, we describe the input catalog and the observations under the APOGEE ancillary program called M33 Clusters \citep[see][Section B.1]{Beaton2021}. We plot locations of the clusters on a color image of M33 in Figure~\ref{fig:fiber_positions}.

For this work we use the following nomenclature: GCs (GCs, old ($\gtrsim$5 Gyr and metal-poor), intermediate-age clusters (IACs, $\sim$1-3 Gyr and more metal-rich), and young star clusters (YSCs, $\lesssim$1 Gyr characterized by broad Brackett absorption features). We intentionally avoid mass- or structure-based terms (e.g., ‘open’ or ‘young massive cluster’) due to the lack of reliable mass estimates. Note that we do not present measured ages for any clusters in this work, and our age ranges here are inferred from visual inspection of the spectra and morphology of the clusters. See Figure~\ref{fig:photspeccomp} for a comparison of spectra and imagery of the three categories.

\subsection{Observations}

The M33 cluster observations were taken as an ancillary program of the Apache Point Observatory Galactic Evolution Experiment \citep[APOGEE,][]{Majewski2017, Beaton2021}. APOGEE, part of the Sloan Digital Sky Survey \citep[SDSS-III and -IV;][]{Eisenstein2011,Blanton2017}, took high-resolution ($R\sim22,500$), NIR ($1.5 - 1.7 \mu$m) spectra of $\sim$650,000 stars in the MW. Additionally it observed individual stars in the Magellenic Clouds and other Local Group (LG) dwarf galaxies, as well as the integrated light of GCs in M33 and M31 \citep{Zasowski2013, Zasowski2017, Beaton2021, Santana2021}.

APOGEE made use of the 2.5-meter Sloan Telescope \citep{Gunn2006} at Apache Point Observatory in New Mexico and the 2.5-meter du~Pont Telescope \citep{BowenVaughan1973} at Las Campanas Observatory in Chile. Both telescopes are connected to an APOGEE spectrograph \citep{Wilson2019} and the data products are processed as described in \citet{Nidever2015}. Stellar parameters and abundances are derived using the APOGEE Stellar Parameters and Chemical Abundances Pipeline \citep[ASPCAP;][]{ASPCAP}. The final data release (DR17) of SDSS-IV, which includes our M33 GC observations, is outlined in \citet{SDSSDR17}. Note that these pipelines are optimized for observations of individual stars, and as such we had to rework the visit stacking and abundance measurement procedures for this analysis. See Section~\ref{sec:dataproc} for specifics.

APOGEE observes in the $H$-band, which contains spectral features from a number of $\alpha$ elements. These elements, such as O, Mg, Si, and Ti, are primarily produced in core collapse supernovae. APOGEE also measures many other elements, such as Fe, Ni, and Mn, which are produced largely in Type Ia supernovae. Such metals, along with $\alpha$ elements, trace the chemical evolution and star formation history of a stellar population \citep[e.g.,][]{Weinberg2017}. Additionally, RGB and AGB stars are brightest in this band, and, due to their long lifetimes are very numerous. The $H$-band is much less affected by dust attenuation than the optical, so APOGEE is able to clearly observe stars throughout most of the galaxy and beyond. Finally, APOGEE's high spectral resolution allows for precise measurements of stellar parameters, abundances, and velocity dispersions, as well as efficient subtraction of skylines and tellurics.

\subsection{Target Selection}

As described in \citet{Beaton2021}, the targets for this study were selected from the extended source catalog of \citet{Fan2014}. This catalog, which is based in part on the catalog of \citet{SanRoman2010} (see Table~\ref{tab:catalogs}), presented homogeneous UVBRI photometry of 708 M33 star cluster candidates based on archival images from the Local Group Galaxies Survey. We transformed their V-band magnitudes to H-band using the strategy of \citet[][see Table 1]{Sakari2016}. We selected all extended sources within a magnitude range, 12.5 $<$ H $<$ 15, which yielded a total of 132 sources. 

To obtain the observations, four plug-plates were designed for the M33 field. Because of the spatial restrictions on how fibers can be plugged, only 95 of our 132 clusters were observed\footnote{These 95 are identifiable in the allStar or allVisit files as spectra in the M33 field whose APOGEE\_ID begin with `AP' rather than `2M'.}. Each fiber position had several sky fibers associated with it that were placed in different locations on each plate. These locations were typically to the upper left, upper right, lower left, and lower right of the fiber position when spacing allowed. This created the pattern seen in Figure \ref{fig:fiber_positions}. This particular strategy was chosen so that the background light of the galaxy could be subtracted from the spectrum the way the sky flux typically would be. In the end, each fiber position was observed 13 times. 

Our final sample of spectra has 92 cluster candidates, as we remove two spectra that had no visits with S/N \textgreater 1.0, as well as one spectrum that was of the nuclear star cluster. Figure~\ref{fig:fiber_positions} shows the 92 cluster candidates and sky fiber positions, as well as the clusters in our Bona Fide and Gold samples (see Section~\ref{sec:bonafides}).

\subsection{Bona Fide Cluster Selection} \label{sec:bonafides}

To extract true clusters from the spurious extended sources, we first searched for archival \textit{HST} imaging and visually classified the sources, with the majority of the imaging coming from the PHATTER survey \citep[GO\#14610 - PI Dalcanton,][]{PHATTER}. We then used positional cross-matching with six established M33 cluster catalogs to identify sources without \textit{HST} imaging, matching our APOGEE coordinates within $1\arcsec$, corresponding to the APOGEE fiber radius (see Table~\ref{tab:catalogs}). Objects identified as clusters by either method formed the primary basis of our Bona Fide sample.

\begin{table*}
\centering
\setlength{\tabcolsep}{3.5pt}
\begin{tabular}{ccccccccccccc}
\toprule
APOGEE ID          & SR10 ID & RA      & DEC     & CS88 & M98 & SM07     & SR10 & B15          & J22 & \textit{HST} & Brackett? & Gold \\ \midrule
AP01324221+3021141 & 714     & 23.1759 & 30.3539 &      &     &          & 2    &              &     &              &         & Y    \\
AP01324289+3035383 & 722     & 23.1787 & 30.594  & Cl*  &     & Cluster  & 1    &              &     &              &         &      \\
AP01330979+3022353 & 968     & 23.2908 & 30.3765 &      &     &          & 3    &              &     & Y            &         &      \\
AP01331112+3034216 & 986     & 23.2963 & 30.5727 &      &     &          & 1    &              &     &              & Y       &      \\
AP01331924+3027503 & 1103    & 23.3302 & 30.464  &      &     &          & 2    &              &     &              &         & Y    \\
AP01332053+3049015 & 1115    & 23.3355 & 30.8171 &      &     &          & 1    & GC Cand. &     &              &         & Y    \\
AP01332304+3050116 & 1138    & 23.346  & 30.8366 &      &     &          & 1    & GC Cand. &     &              &         & Y    \\
AP01332307+3033004 & 1139    & 23.3461 & 30.5501 &      & Y   & Cluster  & 2    & Y            & Y   & Y            &         &      \\
AP01332839+3036229 & 1199    & 23.3683 & 30.6064 & GlC  &     & Cluster  & 3    & Y            & Y   & Y            &         & Y    \\
AP01332868+3041349 & 1203    & 23.3695 & 30.693  & GlC  &     & Cluster  & 3    & GC           & Y   & Y            &         & Y    \\
AP01333136+3040204 & 1239    & 23.3807 & 30.6723 &      &     & Cluster  & 3    & Y            & Y   & Y            &         &      \\
AP01333657+3022079 & 1295    & 23.4024 & 30.3689 &      &     &          & 1    &              &     & Y            & Y       &      \\
AP01333721+3034139 & 1311    & 23.405  & 30.5705 &      & Y   & Cluster  & 2    &              & Y   & Y            &         & Y    \\
AP01333726+3031110 & 1312    & 23.4053 & 30.5197 &      &     &          & 1    &              &     & Y            & Y       &      \\
AP01333777+3057132 & 1322    & 23.4074 & 30.9537 &      &     &          & 1    & GC Cand. &     &              &         & Y    \\
AP01334831+3042420 & 1503    & 23.4513 & 30.7117 &      &     &          & 1    &              &     & Y            & Y       &      \\
AP01335143+3052529 & 1557    & 23.4643 & 30.8814 &      &     &          & 1    &              &     &              & Y       &      \\
AP01335212+3029036 & 1566    & 23.4672 & 30.4843 & GlC  & Y   &          & 3    & GC           & Y   & Y            &         & Y    \\
AP01335512+3047580 & 1601    & 23.4797 & 30.7994 & GlC  & Y   & Cluster  & 3    &              & Y   & Y            &         & Y    \\
AP01335543+3100091 & 1605    & 23.481  & 31.0025 &      &     &          & 1    &              &     &              & Y       &      \\
AP01340021+3037468 & 1672    & 23.5009 & 30.6297 &      &     & Unknown  & 3    & Y            & Y   & Y            & Y       & Y    \\
AP01340197+3039377 & 1704    & 23.5082 & 30.6605 &      & Y   & Cluster  & 3    &              & Y   & Y            &         & Y    \\
AP01340286+3043207 & 1717    & 23.5119 & 30.7224 &      &     & Cluster  & 2    &              & Y   & Y            &         & Y    \\
AP01340384+3047286 & 1724    & 23.516  & 30.7913 &      &     & Cluster  & 3    &              & Y   & Y            &         &      \\
AP01340801+3038380 & 1765    & 23.5334 & 30.6439 & GlC  & Y   & Cluster  & 3    & GC           & Y   & Y            &         & Y    \\
AP01341007+3045292 & 1802    & 23.542  & 30.7581 &      & Y   & Cluster  & 1    &              & Y   & Y            &         &      \\
AP01341109+3047349 & 1822    & 23.5462 & 30.793  &      &     &          & 1    &              & Y   & Y            &         & Y    \\
AP01341152+3034524 & 1829    & 23.548  & 30.5812 &      & Y   & Unknown  & 2    &              & Y   & Y            &         & Y    \\
AP01341501+3041190 & 1896    & 23.5625 & 30.6886 &      &     & Cluster  & 3    & Y            & Y   & Y            &         &      \\
AP01341589+3033026 & 1911    & 23.5662 & 30.5507 &      &     &          & 2    &              & Y   & Y            &         &      \\
AP01342080+3057092 & 1965    & 23.5867 & 30.9526 &      &     &          & 1    & GC Cand. &     &              &         & Y    \\
AP01342092+3030397 & 1970    & 23.5872 & 30.511  &      &     &          & 1    &              &     & Y            & Y       &      \\
AP01342236+3048355 & 1988    & 23.5932 & 30.8099 &      &     &          & 1    &              & Y   & Y            & Y       &      \\
AP01342333+3101137 & 2001    & 23.5972 & 31.0205 &      &     &          & 1    & GC Cand. &     &              &         & Y    \\
AP01343022+3038127 & 2075    & 23.6259 & 30.6369 & GlC  &     & Cluster  & 3    & GC           &     & Y            &         & Y    \\
AP01343813+3029585 & 2145    & 23.6589 & 30.4996 &      &     &          & 1    & GC Cand. &     &              &         & Y    \\
AP01344416+3052190 & 2222    & 23.684  & 30.8719 &      &     & Cluster  & 3    &              &     & Y            &         &      \\
AP01344557+3044229 & 2232    & 23.6899 & 30.7397 &      &     & Chip Gap & 1    &              &     &              &         &      \\ \bottomrule
\end{tabular}
\caption{Information for the selected clusters in M33. The next six columns are classifications from a selection of catalogs of star clusters in M33 |- CS82: \citet{CS1982}, M98: \citet{Mochejska1998}, SM07: \citet{Sarajedini2007}, SR10: \citet{SanRoman2010}, B15: \citet{beasley+2015}, and J22: \citet{Johnson2022}. We provide the APOGEE ID, SR10 ID, RA, and DEC (in degrees) for each cluster in our bona fide sample. A `Y' indicates the cluster is included but that no classification is given in the catalog, whereas empty cells indicate the cluster is not found in that catalog. For M98, classifications are given on a scale from A to D, where A is a highly likely GC, and D is probably not a GC. For SM07, the classification of the object is given. For SR10, classifications are given on a scale from 1 to 3, where 1 is a cluster candidate, 2 is a highly probable cluster candidate, and 3 is a confirmed cluster; the few clusters that are in SR10 but no others are included because their measured radial velocity is \textgreater -100 \kms. The final three columns are \textit{HST}: whether archival imagery of the cluster exists from \textit{HST}, Brackett?: whether strong Brackett absorption features in the spectrum indicating a very young ($\sim$100 Myr) cluster, and Gold: whether the cluster is in our "Gold" sample, see Section~\ref{sec:bonafides}.}
\label{tab:catalogs}
\end{table*}

Our Bona Fide sample consists of 37 clusters across the three broad classes | YSCs, IACs, and GCs, as defined in Section~\ref{sec:sample}. We further narrowed down our sample to the include just the GCs and IACs in our catalog that are of sufficiently high empirical signal-to-noise to be reasonably fit (eS/N $\gtrsim$ 15). This removes all MW foreground stars, as well as the YSCs, for which we do not have adequate models to measure their abundances. However, we still use their measured radial velocities for kinematic comparison to the GCs and IACs.

This final sample of 20 GCs and IACs is called our `Gold' sample (see Table~\ref{tab:catalogs}). We present radial velocity measurements for all clusters in the Bona Fide sample, and velocity dispersion, metallicity, and $\alpha$-abundance measurements for the clusters in the Gold sample. 

\begin{figure*}
\centering
\includegraphics[width=0.8\textwidth]{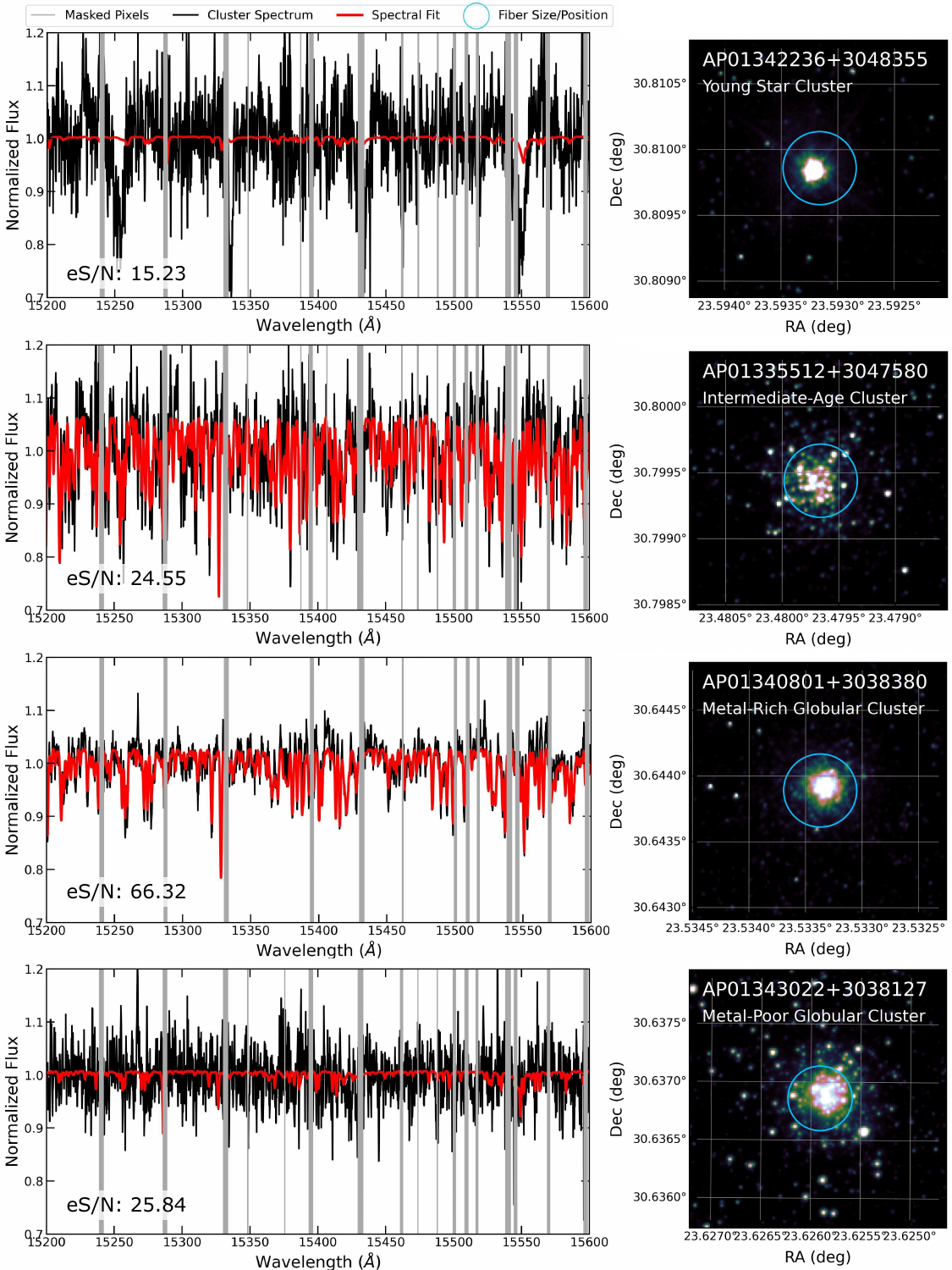}
\caption{Comparison of the APOGEE spectra and \textit{HST} imagery of four clusters in M33. We show the spectrum in black, the best fit to the data in red, and the 2\arcsec\space APOGEE fiber diameter in blue. The four clusters that were selected are (from top to bottom) AP01342236+3048355, a YSC with broad Brackett absorption features, such as the ones near 15250 and 15550\AA; AP01335512+3047580, an IAC with a very clumpy morphology; AP01340801+3038380, a metal-rich GC ([M/H]$\sim$-0.8); and AP01343022+3038127, a metal-poor GC ([M/H]$\sim$-1.8). All four clusters were identified in \citet{Johnson2022}. Imagery for the top three comes from the PHATTER survey \citep{PHATTER} F475W data, with the bottom row coming from GO \#16274 (PI: Sohn), F814W filter. \textsc{Note:} The spectral fit in the YSCs (top row) is only used for radial velocity measurement, no other kinematics or abundances are obtained.}
\label{fig:photspeccomp}
\end{figure*}

\section{Data Processing} \label{sec:dataproc}

Our M33 plates included fibers for individual MW stars, and so used the standard observing modes for H-band observations. As such, our cluster spectra are often low signal-to-noise ratios (S/N) due to their lower apparent brightnesses.

Additionally, the pipeline that APOGEE uses to process and reduce the data is, naturally, optimized for observations of individual stars. While the majority of this pipeline still works for our integrated light observations, the final visit-stacking step had to be reworked to account for errors that arise primarily from the (often) low-S/N ratios of the observations. Sky subtraction, in particular, is challenging for visits with low SNR \citep{Shetrone2026}.

To address these issues, we developed a custom visit-stacking routine that is described in full in Section~2.2 of \citet{Gibson2023} and updated in Section~3.1 of \citet{Gibson2025}. These works will hereafter referred to as \citetalias{Gibson2023} and \citetalias{Gibson2025}, respectively. 

For this work we do not have custom apStar files of the M33 observations as we had in previous works. Instead, we stack visits from the individual apVisit\footnote{\url{https://data.sdss.org/datamodel/files/APOGEE_REDUX/APRED_VERS/visit/TELESCOPE/FIELD/PLATE_ID/MJD5/apVisit.html}} files, for which there is one FITS file for each individual observation of each target in APOGEE. Unlike apStar files, apVisit files a) have separate HDUs for each chip, b) sample each spectrum at 4096 wavelength points per chip\footnote{The native wavelength array for the APOGEE detector has 8575 pixels spanning the all three chips}, and c) no velocity shifts have been applied, so all spectra are in the observed frame. 

Due to these differences, we present a general overview of the visit stacking routine below.

\begin{enumerate}
    \item Locate the apVisit files for each observation of an APOGEE ID using the allVisit file\footnote{\url{https://data.sdss.org/sas/dr17/apogee/spectro/aspcap/dr17/synspec_rev1/allVisit-dr17-synspec_rev1.fits}}.
    \item Convert the APOGEE pixel level bitmasks\footnote{\url{https://www.sdss4.org/dr17/irspec/apogee-bitmasks/\#APOGEE_PIXMASK:APOGEEbitmaskforindividualpixelsinaspectrum}} for each spectrum into a binary mask array flagging pixels affected by bits 0, 1, 2, 3, 4, 5, 6, 7, 8, 9, 10, and 14. We note these are the same bits masked in all previous works.
    \item Interpolate the object flux, error, and sky flux for each chip onto one array corresponding the standard APOGEE wavelength array where all three chips and the gaps between them combined for 8575 total pixels.
    \item Use \texttt{PyAstronomy.pyasl.helcorr} to determine the barycentric correction at the midpoint of each observation in Julian days\footnote{The Sloan Foundation Telescope at Apache Point Observatory has a latitude and longitude of 32.78036 and -105.82042 degrees and an elevation of 2788 meters. Source: \url{https://www.sdss.org/instruments/}.}
    \item Shift the flux, error, and sky arrays from each visit by the barycentric correction so that all visits are in the heliocentric frame.
    \item Re-split the wavelength, flux, error, sky, and mask arrays into the individual chips.
    \item Calculate the psuedo-continuum of each chip in the visit spectra by taking a 500-pixel running median. 
    \item Identify skylines in the spectra by flagging pixels that are above the 85th percentile in sky flux and above the 90th percentile in the $\Delta$Flux parameter, which is the absolute value of the object flux minus the psuedo-continuum. Additionally, mask pixels between 15890 and 15960 in fibers between 20 and 35 (see Section 2.2.1 of \citetalias{Gibson2023} for justification). 
    \item Recalculate the 500-pixel psuedo-continuum of the spectra, this time ignoring the masked pixels from steps 2 and 8.
    \item For each chip of each visit, calculate the median S/N of all the unmasked pixels. Remove chips with an S/N below one.
    \item For each individual chip, stack all remaining psuedo-continuum normalized visit spectra using the inverse-variance weighted average of the object flux. In the final stacked spectrum, mask all pixels that are masked in at least half of the visits. Take the final sky flux and continuum to be the non-weighted average of the sky flux and continuum per pixel across all visits.
    \item Recombine all chips onto the standard APOGEE wavelength array. Note in \citetalias{Gibson2023} and \citetalias{Gibson2025} the three chips were analyzed separately. This change is due to the fitting routine in this work utilizing \texttt{pPXF}, rather than the custom spectral analysis method used in previous works.
\end{enumerate}

In summary, this routine allows us to take a single APOGEE ID and combine all its visits into a single masked, psuedo-continuum normalized, and co-added spectrum. In the end there is one array each for the combined object flux, object flux error, sky flux, masks, continuum, and wavelength corresponding to each pixel.

\section{SSP Models} \label{sec:models}

\begin{figure}
\includegraphics[width=\columnwidth]{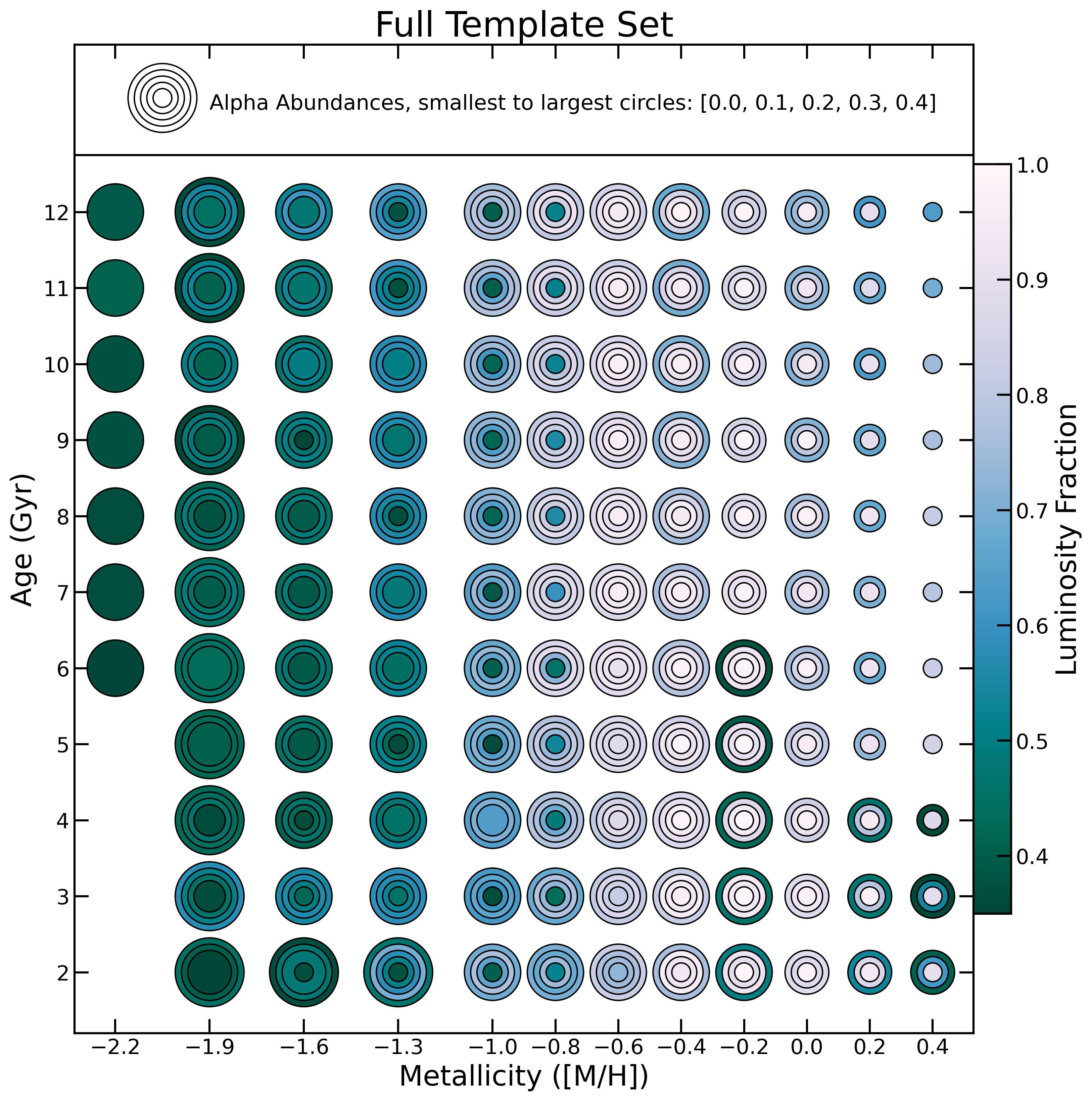}
\caption{The A-LIST sample that has been fed into pPXF, colored by luminosity fraction, intended to demonstrate that our template library well-samples the range of ages, metallicities, and $\alpha$ abundances we might find for the clusters. The concentric circles represent the $\alpha$ abundance of the templates available at each combination of age and metallicity, with smaller circles being lower values of [$\alpha$/M].}
\label{fig:alist_sample}
\end{figure}

We analyzed the data using the APOGEE Library of Infrared SSP Templates \citep[A-LIST\footnote{\url{https://github.com/aishashok/ALIST-library}};][]{Ashok2021}, which is a grid of spectral models created empirically from spectra of MW stars observed by APOGEE. These models were made specifically to analyze integrated light spectra taken by APOGEE and were used in \citetalias{Gibson2023} and \citetalias{Gibson2025}. A-LIST models span a range in Age from 2 to 12 Gyr, Metallicity ([M/H]) from -2.2 to +0.4 dex, and $\alpha$ Abundance ([$\alpha$/M]) from -0.2 to +0.4 dex. 

We make cuts to the model grid based on model reliability and parameter space coverage. For model reliability we use the luminosity fraction parameter (see Section~4.1.1 of \citet{Ashok2021}, which is the ratio of the number of real APOGEE stars used to create a given model spectrum to the total number of stars that would be theoretically expected to exist in the model's age, metallicity, effective temperature, and surface gravity bin. 

Per \citet{Ashok2021}, models with luminosity fraction \textgreater 0.32 should be removed. We have upped this to 0.35 for this work. Additionally we have removed all templates with [$\alpha$/M] \lessthan -0.1, as well as several young (Age \lessthan 5 Gyr), metal poor ([M/H] = -2.2) templates which had luminosity fractions between 0.35 and 0.4. These choices were made primarily to smooth out the shape of the parameter space for better \texttt{pPXF} regularization (See Section~\ref{sec:fitting}), and because such low metallicities at that age are astrophysically unlikely and difficult to measure.

In Appendix~\ref{app:ages} we demonstrate that, due to the minimal variations between models of different ages, our abundance estimates are largely insensitive to the age measured. As such we only present approximate, inferred ages for the three cluster classes.

The final sample of A-LIST templates is visualized in Figure~\ref{fig:alist_sample}.

\section{Spectral Fitting} \label{sec:fitting}

To fit our spectra, we utilize the \textsc{ppxf} software package \citep[Penalized PiXel Fitting, ][]{pPXF_1, pPXF_2, pPXF_3} to fit the A-LIST models to our data.

\subsection{Preparing Spectra for Fitting}

As in previous works, before fitting the spectra, we mask out pixels with untrustworthy fluxes and properly scale the error arrays as in Section~4.2 of \citetalias{Gibson2023} and Section~3.1 of \citetalias{Gibson2025}, with some slight modifications.

For the masking, given the smaller radial velocity of M33 relative to M31, we now only mask out the 100 and 25 pixels on the red and blue ends of each chip, respectively. Furthermore, given the higher signal-to-noise and lower velocity dispersion of these cluster spectra (relative to the M31 disk spectra of \citetalias{Gibson2023} and \citetalias{Gibson2025}), we do not see the same consistently misfit regions in the blue and green chips, so we do not mask those out. The skyline masking remains unchanged. Finally, in this work we take no steps to alter the continuum of our spectra, as the 500-pixel running median psuedo-continuum we use here is adequate for our cluster spectra and necessary for the proper physical interpretation of the results from our \textsc{ppxf} fits (see Section~\ref{sec:abunds}.

Additionally, we calculate the empirical signal-to-noise ratio (eS/N) of each of our spectra. This process is largely the same as in previous works (see Step 2 in Section~4.2 of \citetalias{Gibson2023} and Section~3.1 of \citetalias{Gibson2025}), where we fit a sample of A-LIST templates to our data with \textsc{ppxf} using an \texttt{mdegree} of 40 to obtain as artificially perfect of a fit as possible. For this work, we use a smaller subset of A-LIST templates, restricting the sample to have Luminosity Fraction \greaterthan 0.32, [M/H] either $\leq$ -1.0 or equal to -0.6, -0.2, or 0.2, [$\alpha$/M] = 0.0 or 0.3, and Age = 2.0, 4.0, 8.0, or 12.0 Gyr. This sample contains 48 templates well sampled across the parameter space. We use this fit to scale the errors of our spectra by the size of the residuals (Equation~2 of \citetalias{Gibson2025}) such that the $\chi^2$ of this `perfect' fit is unity, and to get reasonable values of the likelihood and $\chi^2$ of future fits described in the following sections.

\subsection{Determining Radial Velocities} \label{sec:vels}

In previous works we were able to make use of radial velocity measurements from the literature when analyzing our spectra. Unfortunately, such measurements do not exist for many of our clusters, so we had to do a special Markov-Chain Monte Carlo (MCMC) to get accurate kinematics for our clusters before fitting for the abundances and ages. To do this, we perform a two-dimensional MCMC with \texttt{emcee} \citep{emcee}, similar to those done in \citetalias{Gibson2023}, where the only parameters are $V$ and $\sigma_{\rm int}$. We fit each cluster spectrum using a single A-LIST template to reduce the parameter space and simplify the fit and fidelity of the radial velocity measurements. To account for real variations in the stellar population in each spectrum, we first fit with an old, metal-poor template ([M/H] = -1.0, [$\alpha$/M] = +0.3, and Age = 12 Gyr), then a young, metal-rich template ([M/H] = 0.0, [$\alpha$/M] = 0.0, and Age = 2 Gyr). From these two results, we take the radial velocity of the fit with the lowest $\chi^2$.

Due to the low S/N and metallicity of many of our spectra, results for this fit often tended towards fitting the flat continuum of the spectra, as the real spectral features are harder to pick out in the noise. To address this, we imposed an informative prior on our MCMC to push the dispersion of the fit down to physically-plausible values. Typical Milky Way GCs have a velocity dispersion of $\sim$1 km/s  \citep{Baumgardt+Hilker2018}. To enforce a low velocity dispersion in our fits, we added a Gaussian prior to our likelihood with a mean ($\mu$) of 2 \kmspace and standard deviation ($\sigma_{\rm int}^{\prime}$) of 0.5 \kms. It is worth noting that, since they have been detected at a distance, the clusters in our sample are likely more massive than a typical MW GC, and therefore have somewhat higher velocity dispersions. Still, this prior greatly improved the fits to produce results with robust radial velocity determinations. 

With the addition of this prior, our likelihood equation becomes:
\begin{equation}
    L = -\frac{\chi^2}{2} - \frac{(\sigma_{\rm int} - \mu)^2}{2 \sigma_{\rm int}^{\prime 2}} \label{eq:likelihood}
\end{equation}

We note that convergence is not needed for this MCMC, as we use it simply to locate the region of maximum likelihood, and do not need to fully characterize, e.g., the size of the region. We use this value as the starting guess for the pPXF fitting routine described in the following section. For this MCMC fit, we used 1000 walkers and 150 iterations. We initialized the walkers with -250 $\leq V \leq$ -100 \kmspace \citep[centered around M33's systemic radial velocity of -179 \kms,][]{Corbelli+Salucci2000}, and 3 $\leq \sigma_{\rm int} \leq$ 10 \kms. We allowed the walkers to explore values from -500 $\leq V \leq$ +100 \kmspace and 0.1 $\leq \sigma_{\rm int} \leq$ 30 \kms. The result from this fit is the set of parameters that produce the maximum likelihood. 

For future fits, we allow the radial velocity to vary in a small range around this measured value ($\pm$5 \kms), but let the dispersion vary between 0.1 and 20 \kms, as the best fit value will change slightly with metallicity.

\subsection{Measuring Chemical Abundances} \label{sec:abunds}

Now that we have a reliable measurement of each cluster's radial velocity, we can employ \textsc{ppxf} to measure their light-weighted average metallicity and $\alpha$ abundance. \textsc{ppxf} fits each spectrum as a linear combination of the A-LIST, allowing the model to interpolate between the discrete templates in the library. The resulting template weights are used to characterize the light-weighted stellar population, and should not be interpreted as evidence for multiple distinct stellar populations \citep[see Section~3.5 of][]{pPXF_2}.

For these fits, we set both \texttt{degree} and \texttt{mdegree} equal to -1 so that no additive or multiplicative polynomials are applied to the fit. With this, our A-LIST templates and cluster spectra are continuum normalized the exact same way, and so we can take the average of the templates used in the fits, weighted by the weights calculated by \textsc{ppxf}, to be the light-weighted [M/H] and [$\alpha$/M] of each spectrum.

We also apply a regularization of 20 to these fits to smooth out the distribution of templates used by \textsc{ppxf}. We determined 20 to be the value at which, for a range of eS/N, the non-reduced $\chi^2$ of the regularized fit increased from the unregularized value by $\sqrt{2\times\textrm{DOF}}$, where DOF is the number of unmasked pixels in the spectrum\footnote{This test is outlined in the \textsc{ppxf} documentation, \url{https://pypi.org/project/ppxf/}, under the \texttt{regul} parameter}. Typically, for model spectra such as A-LIST that vary in three dimensions ([M/H], [$\alpha$/M], and Age), one would sort the models into a four-dimensional array of shape (number of pixels in a template, number of template ages, number of template metallicities, number of template $\alpha$ abundances). However, given that our template coverage is not a perfect 4-D cube (see Figure~\ref{fig:alist_sample}), we instead input our models as a 2-D array of shape (number of pixels in a template, number of templates), but sort them such that the changes between adjacent templates are both minimal and informative\footnote{Since the changes in metallicity are most significant, we first want all of the templates of a given metallicity next to each other. Since the changes in age are minimal, we want all the templates of a given age at a certain metallicity next to each other. This ensures that the most significant change from one template to the next is the $\alpha$ abundance, therefore ensuring that the shift from one template to the next is informative of this parameter. The resulting order of templates would be something like: (Age, [M/H], [$\alpha$/M]) = (12.0, -1.0, 0.1), (12.0, -1.0, 0.0), (11.0, -1.0, 0.0), (11.0, -1.0, 0.1), (11.0, -1.0, 0.2), (11.0, -0.8, 0.2), (11.0, -0.8, 0.1), and so on.}.

We perform our fits with \texttt{moments}=2, to measure the velocity, $v$, and velocity dispersion, $\sigma_{\rm int}$ of each spectrum. We set the initial guess for $v$ to the value measured in Section~\ref{sec:vels} and bound the fit to be within 5 \kmspace of this value. We set the initial guess for $\sigma_{\rm int}$ to be the dispersion result from \ref{sec:vels}, but allow the fit to vary anywhere between 0.1 and 20 \kms. 

To measure observational uncertainties on the fits, we utilize the jackknife resampling method of \citet{jackknife}. In this method, each visit is systematically removed from the sample and the remaining visits are combined and analyzed. Each fit is performed the same way using the methodology described above. This method then gives us $N$ different measurements for the best fit parameters for a given cluster ($N$ is usually 10 to 13, depending on the number of visits to a given fiber position with S/N \textgreater 1.0). We take the final measurement of each cluster's kinematics and abundances to be the average values of the jackknife-resampled measurements, $x_{jack}$, and the errors to be jackknife standard error: 

\begin{equation}
x_{err} = \sqrt{\frac{(N-1)}{N}\sum_{i=0}^N(x_i-x_{jack})^2}.
\end{equation}

\section{Results} \label{sec:results}

Our spectral analysis has produced measurements of the kinematics and chemistry of the clusters. In Section~\ref{sec:kinematics} we present the radial velocities of the Bona Fide sample clusters and compare them to the radial velocity of the disk. In Section~\ref{sec:populations} we present the chemical abundances of the 20 Gold sample clusters and characterize populations of clusters — our results are given in Table~\ref{tab:results}. The Gold sample is our highest fidelity clusters | that is, they are the spectra that we are most confident are clusters, and that have high enough eS/N to have precision chemical abundance measurements. Note that we do not present measured ages for our clusters (see Appendix~\ref{app:ages}), and any age ranges given are qualitatively inferred from spectral inspection and literature comparison.

\subsection{Kinematics of the M33 Star Clusters} \label{sec:kinematics}

\begin{figure}
\centering
\includegraphics[width=\columnwidth]{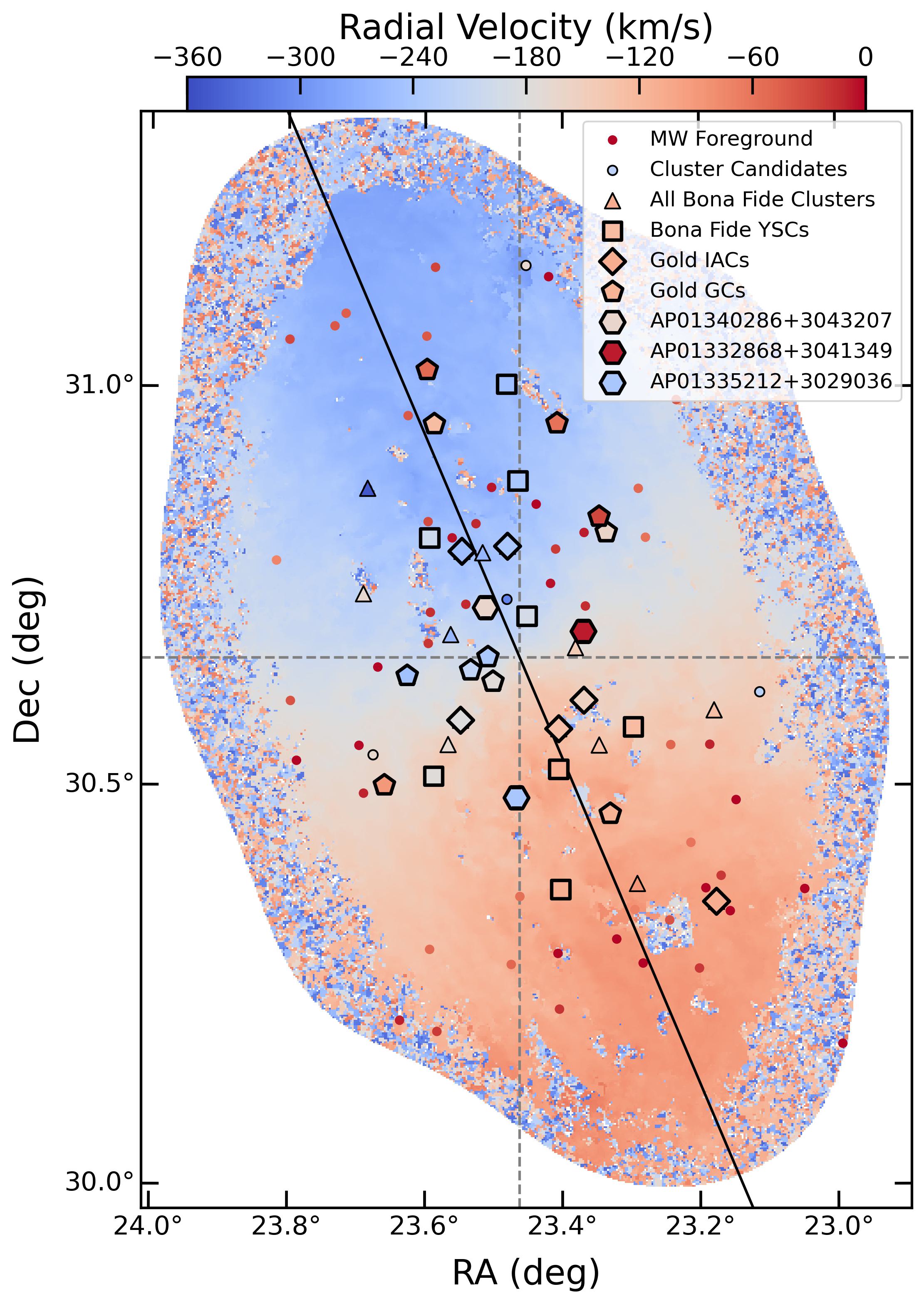}
\caption{Radial velocity of all 93 spectra in our sample, plotted over the H\textsc{I} radial velocity map from \citet{Gratier2010}. The Gold sample GCs and IACs are shown as pentagons and diamonds, respectively, while the YSCs are the squares. All remaining Bona Fide sample clusters are shown as triangles. The three hexagons are our likeliest candidates for accretion as described in Section~\ref{sec:oddballs}. The cluster candidates (see Section~\ref{sec:kinematics}) and MW foreground stars are shown as small dots with and without a border.}
\label{fig:cluster_vels}
\end{figure}

In Figure~\ref{fig:cluster_vels} we plot the radial velocity of all of our spectra overtop the H\textsc{I} gas velocity map of M33, highlighting the Bona Fide and Gold sample clusters. We present radial velocity ($RV$) measurements for all 37 of our Bona Fide clusters in Table~\ref{tab:vels}, and discuss the results below.

\begin{table}
\centering
\begin{tabular}{ccccc}
\toprule
ID                 & V       & $\rm V_{err}$ & $\rm V_{H\textsc{I}}$   & Class \\ \midrule
AP01324221+3021141 & -109.12 & 0.65   & -100.08 & IAC   \\
AP01324289+3035383 & -126.72 & 0.63   & -150.05 &       \\
AP01330979+3022353 & -88.53  & 1.58   & -98.13  &       \\
AP01331112+3034216 & -123.75 & 5.24   & -134.25 & YSC   \\
AP01331924+3027503 & -115.08 & 2.06   & -118.79 & GC    \\
AP01332053+3049015 & -160.2  & 0.43   & -208.59 & GC    \\
AP01332304+3050116 & -28.92  & 0.61   & -210.02 & GC    \\
AP01332307+3033004 & -129.56 & 0.28   & -137.32 &       \\
AP01332839+3036229 & -152.14 & 0.31   & -151.73 & IAC   \\
AP01332868+3041349 & -7.96   & 0.5    & -184.2  & GC    \\
AP01333136+3040204 & -143.57 & 3.49   & -180.3  &       \\
AP01333657+3022079 & -110.36 & 2.84   & -106.12 & YSC   \\
AP01333721+3034139 & -129.9  & 0.22   & -129.12 & IAC   \\
AP01333726+3031110 & -111.54 & 1.1    & -113.14 & YSC   \\
AP01333777+3057132 & -57.98  & 0.54   & -243    & GC    \\
AP01334831+3042420 & -191.41 & 0.55   & -206.78 & YSC   \\
AP01335143+3052529 & -239.95 & 4.35   & -243.01 & YSC   \\
AP01335212+3029036 & -240.78 & 0.25   & -122.56 & GC    \\
AP01335512+3047580 & -238.02 & 0.2    & -239.69 & IAC   \\
AP01335543+3100091 & -251.5  & 1.93   & -253.76 & YSC   \\
AP01340021+3037468 & -175.32 & 0.26   & -178.46 & GC    \\
AP01340197+3039377 & -238.79 & 0.2    & -193.98 & GC    \\
AP01340286+3043207 & -162.17 & 0.5    & -220.34 & GC    \\
AP01340384+3047286 & -261.36 & 0.64   & -249.82 &       \\
AP01340801+3038380 & -215.95 & 0.09   & -187.08 & GC    \\
AP01341109+3047349 & -252.06 & 0.33   & -245.32 & IAC   \\
AP01341152+3034524 & -179.89 & 0.17   & -166.93 & IAC   \\
AP01341501+3041190 & -260.08 & 10.32  & -199.76 &       \\
AP01341589+3033026 & -168.21 & 0.33   & -142.14 &       \\
AP01342080+3057092 & -125.91 & 0.99   & -267.33 & GC    \\
AP01342092+3030397 & -175.94 & 3.98   & -147.91 & YSC   \\
AP01342236+3048355 & -201.44 & 8.22   & -243.06 & YSC   \\
AP01342333+3101137 & -53.14  & 0.36   & -264.51 & GC    \\
AP01343022+3038127 & -247.94 & 0.48   & -191.34 & GC    \\
AP01343813+3029585 & -90     & 0.28   & -160.95 & GC    \\
AP01344416+3052190 & -349.71 & 27.48  & -254.33 &       \\
AP01344557+3044229 & -166.41 & 0.63   & -216.82 &       \\ \bottomrule
\end{tabular}
\caption{Bona Fide sample cluster radial velocities and the disk H\textsc{I} radial velocity at each cluster's position from \citep{Gratier2010}. We also include classifications for the YSCs, as well as GCs and IACs that are in our Gold sample. Clusters with missing classifications do not have sufficience eS/N to be robustly classified.}
\label{tab:vels}
\end{table}

The 37 clusters in our Bona Fide sample span radial velocities from approximately $-350$ to $-10$~\kms, with a median uncertainty of just $0.4$~\kms. Many closely follow the rotation of the M33 H\textsc{I} disk (Figure~\ref{fig:cluster_vels}), although several notable outliers are present, and are discussed further in Section~\ref{sec:oddballs}.

The velocity distribution is centered near the systemic velocity of M33 ($\sim-180$~\kms), but contains a clear excess of clusters with less blueshifted velocities than M33, which could be indicative of accreted origin (see Section~\ref{sec:retrograde}. The most blueshifted object in our sample belongs to the young star cluster population and is located in the outer disk at a galactocentric radius exceeding 3 kpc.

The Gold sample clusters have internal velocity dispersions ranging from approximately $1$ to $10$~\kms, with a median value of $2.8$~\kms and a median uncertainty of $0.75$~\kms. These values are typical of massive GCs and compact star clusters \citep[e.g.,][]{Baumgardt+Hilker2018}.

We present the radial velocities of the Bona Fide sample, our classifications into YSC, IAC, or GC, and the H\textsc{I} velocity of the disk at each GC location in Table~\ref{tab:vels}

\begin{figure*}
\centering
\includegraphics[width=0.8\textwidth]{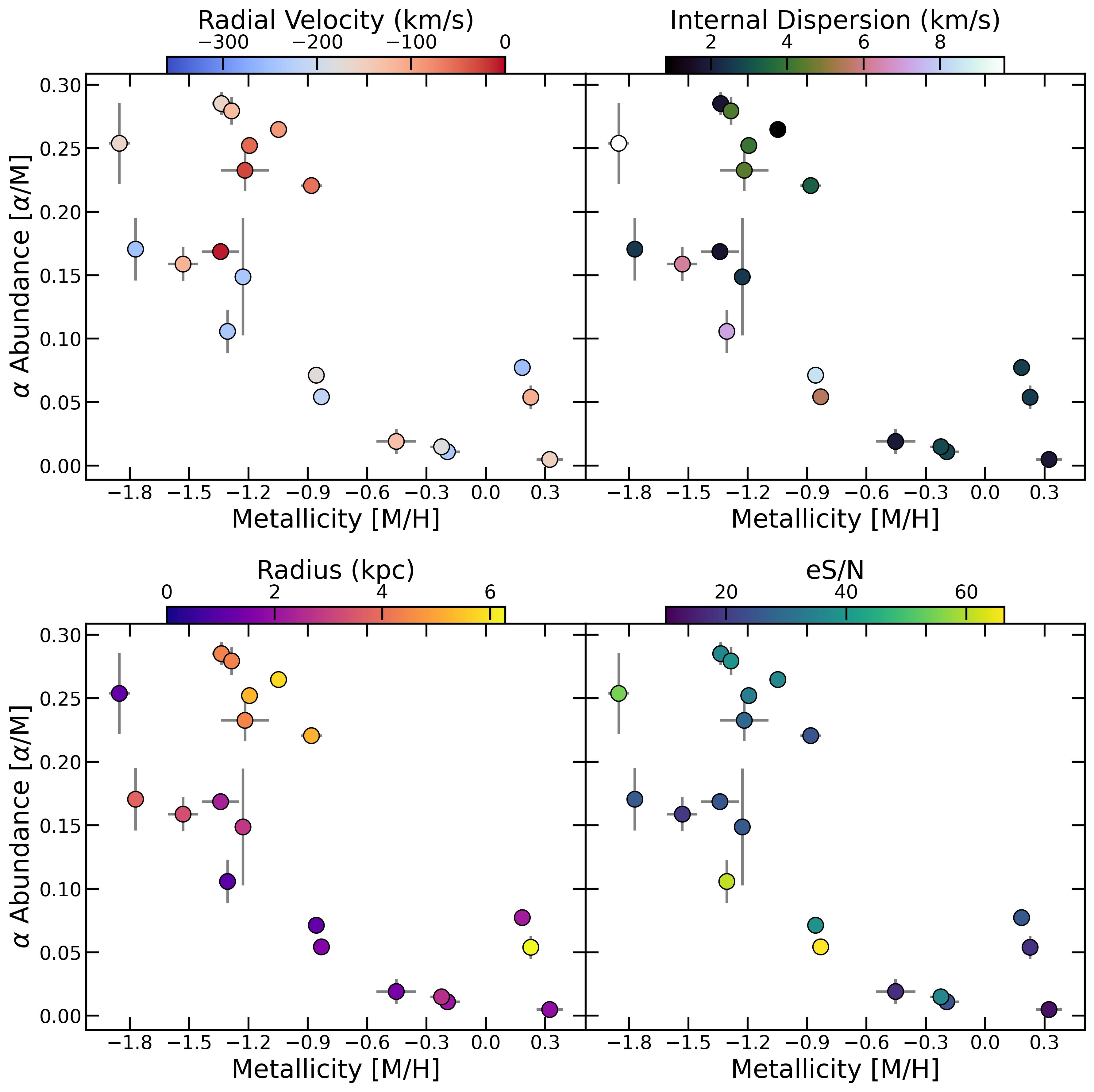}
\caption{Results of our spectral analysis of the Gold sample clusters, plotted in the [$\alpha$/M] vs. [M/H] plane. The upper left panel shows clusters colored by radial velocity, the upper right shows clusters colored by velocity dispersion, the lower left shows clusters colored by their galactocentric radius, deprojected into the plane of the disk, and the lower right shows clusters colored by eS/N. We assume a position angle of 23\degrees and an inclination of 54\degrees \citep{McConnachie2006}.}
\label{fig:cluster_results}
\end{figure*}

Figure~\ref{fig:cluster_vels} includes the radial velocities of the remaining, non-confirmed cluster candidates (small points). Most have velocities consistent with Milky Way foreground stars. Four candidates, however, have velocities consistent with M33 and merit further investigation: AP01344211+3032158 ($RV\approx-170$~\kms) in the eastern outer disk, AP01334865+3109054 ($RV\approx-165$~\kms) in the northern halo, AP01322693+3036594 ($RV\approx-222$~\kms) in the western halo, and AP01335532+3043580 ($RV\approx-313$~\kms) in the inner disk. The first three exhibit integrated spectra resembling those of confirmed clusters, while AP01322693+3036594 also has the highest measured $\alpha$ abundance in our full sample despite its relatively low eS/N. AP01335532+3043580 instead shows possible Brackett absorption features suggestive of a young stellar population.

No high-resolution, space-based imagery of these objects exists yet, but we present CFHT MegaCam images from PAndAS \citep{McConnachie2006} of these four candidates in Figure~\ref{fig:other_pandas}. Of the four, only AP01335532+3043580 appears unambiguously extended or amorphous. These objects have been included in previous ground-based cluster catalogs \citep[e.g.,][]{SanRoman2010}, so their unusual radial velocities make them compelling targets for future high-resolution imaging and spectroscopy.

\subsection{Populations identified in [M/H] vs. [$\alpha$/M] plane} \label{sec:populations}

\begin{table*}
\centering
\begin{tabular}{cccccccccccc}
\toprule
ID                 & eS/N  & $\chi^2$ & $R$ & $\rm RV$ & $\rm RV_{err}$ & $\rm \sigma_{\rm int}$ & $\rm \sigma_{err}$ & $\rm [M/H]$ & $\rm [M/H]_{err}$ & $\rm [\alpha/M]$ & $\rm [\alpha/M]_{err}$ \\ \midrule
AP01324221+3021141 & 18.1  & 2.37     & 6.28 & -109.12 & 0.65          & 2.58         & 0.55               & 0.23        & 0.02              & 0.05             & 0.01                   \\
AP01331924+3027503 & 19.44 & 1.84     & 3.32 & -115.08 & 2.06          & 6.25         & 1.31               & -1.53       & 0.08              & 0.16             & 0.01                   \\
AP01332053+3049015 & 36.27 & 1.47     & 4.37 & -160.2  & 0.43          & 1.76         & 1.12               & -1.34       & 0.05              & 0.29             & 0.01                   \\
AP01332304+3050116 & 29.08 & 1.53     & 4.44 & -28.92  & 0.61          & 4.42         & 0.49               & -1.22       & 0.12              & 0.23             & 0.02                   \\
AP01332839+3036229 & 12.56 & 1.63     & 1.9  & -152.14 & 0.31          & 1.8          & 0.9                & 0.32        & 0.07              & 0.0              & 0.0                    \\
AP01332868+3041349 & 25.01 & 1.59     & 2.28 & -7.96   & 0.5           & 1.68         & 1.35               & -1.34       & 0.09              & 0.17             & 0.01                   \\
AP01333721+3034139 & 17.88 & 1.56     & 1.48 & -129.9  & 0.22          & 1.84         & 0.55               & -0.45       & 0.1               & 0.02             & 0.01                   \\
AP01333777+3057132 & 24.7  & 1.55     & 5.2  & -57.98  & 0.54          & 3.34         & 1.08               & -0.88       & 0.05              & 0.22             & 0.01                   \\
AP01335212+3029036 & 25.66 & 1.43     & 2.82 & -240.78 & 0.25          & 2.53         & 0.79               & -1.23       & 0.04              & 0.15             & 0.05                   \\
AP01335512+3047580 & 24.55 & 1.51     & 2.07 & -238.02 & 0.2           & 2.78         & 0.38               & -0.19       & 0.06              & 0.01             & 0.0                    \\
AP01340021+3037468 & 38.24 & 2.91     & 1.13 & -175.32 & 0.26          & 8.44         & 0.41               & -0.86       & 0.03              & 0.07             & 0.01                   \\
AP01340197+3039377 & 60.96 & 1.46     & 1.01 & -238.79 & 0.2           & 7.18         & 0.45               & -1.31       & 0.02              & 0.11             & 0.02                   \\
AP01340286+3043207 & 54.76 & 1.54     & 1.15 & -162.17 & 0.5           & 9.68         & 0.94               & -1.85       & 0.05              & 0.25             & 0.03                   \\
AP01340801+3038380 & 66.32 & 1.71     & 1.68 & -215.95 & 0.09          & 5.56         & 0.3                & -0.83       & 0.02              & 0.05             & 0.01                   \\
AP01341109+3047349 & 25.44 & 1.96     & 2.19 & -252.06 & 0.33          & 2.64         & 0.25               & 0.18        & 0.01              & 0.08             & 0.0                    \\
AP01341152+3034524 & 36.16 & 1.81     & 2.63 & -179.89 & 0.17          & 2.86         & 0.25               & -0.22       & 0.06              & 0.01             & 0.0                    \\
AP01342080+3057092 & 38.84 & 1.51     & 4.36 & -125.91 & 0.99          & 4.31         & 0.87               & -1.28       & 0.04              & 0.28             & 0.01                   \\
AP01342333+3101137 & 33.56 & 1.47     & 5.29 & -53.14  & 0.36          & 3.99         & 0.73               & -1.19       & 0.04              & 0.25             & 0.01                   \\
AP01343022+3038127 & 25.84 & 1.39     & 3.76 & -247.94 & 0.48          & 2.52         & 1.06               & -1.77       & 0.03              & 0.17             & 0.02                   \\
AP01343813+3029585 & 36.47 & 1.53     & 5.79 & -90.0   & 0.28          & 0.83         & 1.35               & -1.05       & 0.04              & 0.26             & 0.01                   \\ \bottomrule
\end{tabular}
\caption{Table of final kinematics (given in \kms) and abundances (given in dex) from the spectral fitting of our Gold sample. Also presented are the eS/N, $\chi^2$ of the fit, and deprojected galactocentric radius, $R$ (given in kpc).}
\label{tab:results}
\end{table*}

The chemical space provides the best plane to distinguish the different populations in our sample. Figure~\ref{fig:cluster_results} shows the radial velocity, internal velocity dispersion, radius, and eS/N of our Gold sample clusters plotted in the $\mathrm{[M/H]}$--$[\alpha/\mathrm{M}]$ plane. We present our kinematic and chemical results for our Gold sample clusters in Table~\ref{tab:results}. 

When combined with positional and radial velocity information, the $\mathrm{[M/H]}$--$[\alpha/\mathrm{M}]$ plane offers the opportunity to look for signs of accretion into M33 (see Section~\ref{sec:oddballs}). Our primary conclusions are drawn from Figures \ref{fig:photspeccomp}, \ref{fig:cluster_results}, \ref{fig:cluster_vels}, and \ref{fig:pop_loc_sep}.

Based on the metallicities and $\alpha$ abundances, we identify three distinct populations in our Gold sample clusters:

\subsubsection{Population One - Globular Clusters}

Based on their chemical abundances and, where available, \textit{HST} imaging, the clusters with $\mathrm{[M/H]}\lesssim-0.5$ are identified as classical GCs. Their integrated spectra are characterized by relatively weak absorption features, with only a small number of narrow metal lines (bottom two rows of Figure~\ref{fig:photspeccomp}). They span a wide range of galactocentric radii and exhibit a range of kinematic behaviors with respect to the H\textsc{I} rotation curve (Figures~\ref{fig:cluster_vels} and \ref{fig:HI_comp}). 

Chemically, it consists of a high-$\alpha$ sequence with $\mathrm{[M/H]}\approx-1.4$ to $-0.8$ centered at $\mathrm{[\alpha/M]}\approx0.25$, and an intermediate-$\alpha$ sequence with $\mathrm{[\alpha/M]}\approx+0.05$ to $+0.17$. It has a maximal $\rm [\alpha/M]$ spread of nearly 0.2 dex at [M/H]=-1.3, and could potentially be bimodal. This is similar to trends seen in the MW's cluster population (see Section~\ref{sec:disk_comp}). 

Our GCs exhibit no statistically significant metallicity gradient. In contrast, the $\alpha$ abundance increases with galactocentric radius, with a slope of $0.043\pm0.007$~\dexkpc. If $\alpha$ abundance traces the star formation timescale of the environment \citep[e.g.,][]{Johnson2021}, this trend suggests that the outer GCs formed from more rapidly enriched material than those at smaller radii. The GCs also become progressively less consistent with the H\textsc{I} disk kinematics as $\alpha$ abundance increases, although the velocity dispersion remains substantial across the full range of abundances.

Taken together, these observations suggest that the metal-poor GCs are not a homogeneous population. Many are broadly consistent with an \textit{in-situ} origin, sharing chemical properties with the M33 disk (Section~\ref{sec:disk_comp}) and exhibiting kinematics compatible with a dynamically heated disk population. However, the combination of bimodal $\alpha$ abundances and increasing kinematic offsets indicates that at least a subset likely has a different evolutionary history.

Particularly noteworthy is a sequence of six clusters spanning $(\mathrm{[M/H]},[\alpha/\mathrm{M}])\approx(-1.4,0.3)$ to $(-0.9,0.2)$, all located at projected galactocentric radii of $\sim5$ kpc. These objects, shown as the large circles in Figure~\ref{fig:cluster_vels}, were all classified as GC candidates by \citet{beasley+2015}. They share remarkably similar chemical abundances and occupy a narrow range of redshifted velocities ($-160$ to -30~\kms), so they are moving toward the MW slower than the rest of the galaxy. Their velocities are inconsistent with the H\textsc{I} rotation curve, particularly on the northern side of the galaxy, suggesting that they may trace a common kinematic structure rather than representing unrelated halo clusters.

Although none of these objects have been observed with \textit{HST}, CFHT MegaCam imaging (Figure~\ref{fig:beasley_pandas}) shows that they are extended and asymmetric, unlike foreground stars in the PAndAS data. While their \textit{Gaia} parallaxes and proper motions are formally non-zero, this behavior is also seen among confirmed M33 clusters and therefore does not argue against cluster membership. 

Beyond this coherent group, several additional clusters warrant individual discussion because their chemistry, kinematics, and spatial locations distinguish them from the broader GC population. AP01332868+3041349, AP01340286+3043207, and AP01335212+3047580 are examined in detail in Section~\ref{sec:oddballs}.

\subsubsection{Population Two - Disky IACs}

These are the metal-rich, $\alpha$-poorest clusters in our sample, with metallicities spanning from $\mathrm{[M/H]}\approx-0.5$ to $+0.3$, and $\alpha$ abundances spanning from $\mathrm{[\alpha/M]}\approx0.0$ to $+0.08$ dex. Their spectra are dominated by narrow metal absorption features with little or no Brackett absorption (see the second row of Figure~\ref{fig:photspeccomp}, consistent with intermediate-age stellar populations dominated by late-F and G-type stars. Existing age estimates place most of these clusters between a few hundred Myr and $\sim$2 Gyr (see Section~\ref{sec:spec_comp}).

These clusters are largely superimposed on the main star forming disk of M33, with radii \lessthan $\sim$3 kpc, and their velocities trace the H\textsc{I} velocity to within $\sim$30 \kmspace (see the triangles in Figure~\ref{fig:cluster_vels} and Figure~\ref{fig:HI_comp}). Our chemistry results are broadly consistent with those of studies of individual stars in the M33 disk \citep[e.g.,][]{Urbaneja2005}, though with a larger spread in abundance. We do not have enough clusters in this sample to robustly measure a metallicity or $\alpha$ abundance gradient, but these clusters' near-solar chemistry, small velocity offsets from the H\textsc{I} disk, and superposition on the star-forming disk all support an \textit{in-situ} origin from chemically enriched disk gas. 

One cluster in this population, AP01324221+3021141 (SR714), shares all of these properties, and yet is the highest-radius cluster in our Gold sample, over 6 kpc from the galaxy's center. This cluster is chemically indistinguishable from the intermediate-age disk cluster population, with $\mathrm{[M/H]}=0.23\pm0.10$ and $\mathrm{[\alpha/M]}=0.05\pm0.01$, and its radial velocity is fully consistent with local disk rotation.

\begin{table*}
\centering
\setlength{\tabcolsep}{3.5pt}
\begin{tabular}{ccccccccc}
\toprule
APOGEE ID           & SR10 ID & This Work {[}M/H{]} & This Work RV & B15 Age & B15 {[}M/H{]} & B15 RV & C02 Age & C02 RV \\ \midrule
AP01332053+3049015  & SR1115  & -1.34               & -160.20      & 9.967   & -1.13         & -160   &         &        \\
AP01332304+3050116  & SR1138  & -1.22               & -28.92       & 0.017   & -1.04         & -28    &         &        \\
AP01332307+3033004  & SR1139  &                     & -129.56      & 1.179   & -1.33         & -127   & 0.032   & -133   \\
AP01332839+3036229  & SR1199  & 0.32                & -152.14      & 0.723   & -0.28         & -132   &         &        \\
AP01332868+3041349  & SR1203  & -1.34               & -7.96        & 5       & -1.63         & -192   &         &        \\
AP01333136+3040204  & SR1239  &                     & -143.57      & 0.92    & -0.25         & -154   & 1.778   & -200   \\
AP01333721+30434139 & SR1311  & -0.45               & -129.90      &         &               &        & 0.016   & -112   \\
AP01333777+3057132  & SR1322  & -0.88               & -57.98       & 11.094  & -0.58         & -62    &         &        \\
AP01335212+3029036  & SR1566  & -1.23               & -240.78      & 9.975   & -1.15         & -215   &         &        \\
AP01335512+3047580  & SR1601  & -0.19               & -238.02      &         &               &        & 0.126   & -228   \\
AP01340021+3037468  & SR1672  & -0.86               & -175.32      & 1.81    & -0.22         & -169   &         &        \\
AP01340801+3038380  & SR1765  & -0.83               & -215.95      & 12.443  & -0.86         & -208   & 5.012   & -218   \\
AP01341152+3034524  & SR1829  & -0.22               & -179.89      &         &               &        & 0.316   & -143   \\
AP01341501+3041190  & SR1896  &                     & -260.08      & 0.161   & -0.02         & -131   &         &        \\
AP01342080+3057092  & SR1965  & -1.28               & -125.91      & 15.724  & -1.28         & -104   &         &        \\
AP01342333+3101137  & SR2001  & -1.19               & -53.14       & 9.733   & -0.86         & -47    &         &        \\
AP01343022+3038127  & SR2075  & -1.77               & -247.94      & 11.74   & -1.74         & -238   & 1.585   & -213   \\
AP01343813+3029585  & SR2145  & -1.05               & -90.00       & 9.977   & -1.01         & -94    &         &        \\
AP01344416+3052190  & SR2222  &                     & -349.71      &         &               &        & 1.585   & -289   \\ \bottomrule
\end{tabular}
\caption{Results from \citet{chandar+2002} (C02) and \citet{beasley+2015} (B15) for clusters in our Bona Fide sample, as well as results from this work for comparison. We show Ages in Gyr and RV in \kmspace. Empty cells for our [M/H] are for bona fide clusters that are not in our Gold sample, and for which we do not present metallicities.}
\label{tab:spec_crossmatch}
\end{table*}

The object has received relatively little attention in the literature. It was first cataloged as a probable cluster by \citet{SanRoman2010}, although other studies have instead identified luminous individual stars within 2\arcsec. No HST imaging is currently available to resolve the stellar population, but we present PAndAS imagery of this cluster and the other two most metal rich IACs in our Gold sample in Figure~\ref{fig:met_rich_pandas}. Additionally, there is evidence of ongoing star formation in M33 out to 10-12 kpc \citep[e.g.,][]{Sharma2011}, so the presence of a young stellar population in this area is plausible. Despite the uncertainty surrounding its detailed nature, its chemistry and kinematics strongly suggest formation from chemically enriched disk gas. Its unusually large galactocentric radius therefore may reflect recent star formation in the outer disk rather than an accreted origin. However, the current observed location is likely to be at a part of the orbit where the cluster spends the most time because its orbit has slowed down. Lots of orbital configurations are possible that might put the cluster in an unlikely position, so without more information, it's hard to say anything.

\subsubsection{Population Three - YSCs}

Population 3 consists of Bona Fide sample clusters whose integrated spectra are dominated by strong Brackett absorption features, with relatively weak metal lines (top row of Figure~\ref{fig:photspeccomp}). These spectra are characteristic of younger stellar populations dominated by A-type stars and are consistent with ages of order $\sim$100 Myr.

Existing \textit{HST} imaging reveals that members of this population are either compact, irregular clusters or loose associations of bright stars, consistent with their young ages. They are primarily located within the actively star-forming disk of M33 and have radial velocities ranging from $-250$ to $-110$~\kms, typically agreeing with the local H\textsc{I} velocity to within $\sim40$~\kms (black dots in Figure~\ref{fig:HI_comp}). Together, their morphology, kinematics, and locations strongly support their identification as recently formed clusters within the M33 disk.

We do not include this population in Figure~\ref{fig:cluster_results}, as these spectra are not well reproduced by our fitting models (see the top row of Figure~\ref{fig:photspeccomp}, which have a minimum age of 2 Gyr and therefore lack young models with significant Brackett absorption. Consequently, we frequently derive unphysical parameters, including artificially high velocity dispersions, low metallicities, and enhanced $\alpha$ abundances. We therefore do not consider the spectroscopic abundances measured for this population to be reliable, and we only present radial velocity measurements.

It is worth noting that these YSCs were analyzed in \citet{Fan2014}, who used integrated-light SED fitting to derive their ages and masses, finding they were between roughly 0.1 and 2 Gyr old between $3\times10^{4}$ and $3\times10^5$ $\rm M_{\odot}$.

\section{Discussion} \label{sec:discussion}

The star cluster population of M33 has been the subject of dozens of studies since \citet{Hiltner1960}. Differences in observational data, survey footprints, cluster selection methods, and catalog construction have produced a highly fragmented literature, with substantial but often incomplete overlap between catalogs. Untangling the relationships among all published samples would require a dedicated meta-analysis in its own right and is therefore beyond the scope of this paper. Instead, we present measurements of our Gold sample clusters from the two most relevant spectroscopic studies, \citet{beasley+2015} and \citet{chandar+2002}, in Table~\ref{tab:spec_crossmatch}. We present a similar table for the three photometric studies with the greatest overlap with our sample, \cite{Sarajedini2007}, \citet{Fan2014}, and \citet{Johnson2022}, in Table~\ref{tab:phot_crossmatch}. In Section~\ref{sec:spec_comp} we compare our results to the broad trends identified in these studies. In Section~\ref{sec:disk_comp} we compare our results to the cluster population of the MW. Then in Section~\ref{sec:retrograde} we assess the orbital sense of the clusters, before detailing three clusters that seem most likely to be accreted in Section~\ref{sec:oddballs}.

\subsection{Comparison of Broader Trends to Previous Spectroscopic Studies of M33 Clusters} \label{sec:spec_comp}

\citet{CS1983} were the first to study the integrated spectra of star clusters in M33. Using measurements of a few indices, namely H$\beta$, MgII, and CN, they identified the correlation between cluster age and metallicity, and that the bluest (youngest, in their estimation) clusters had kinematics that well-traced the H\textsc{I} disk, while the reddest (oldest) clusters deviated from that trend. While we do not present age measurements for our clusters, it is well established that $\alpha$ abundance trends broadly with age in the Milky Way, in that stars at higher-$\alpha$ tend to be older. We find a broad trend between metallicity and $\alpha$ abundance for our clusters: that lower-[M/H] clusters tend to be higher-[$\alpha$/M], and vice versa. In this sense, our results are broadly consistent with the literature.

The later \citet{CS1991} analyzed spectra of 45 clusters. They primarily used the data to measure radial velocities, but were able to split their sample into a young, intermediate, and old classification on the integrated color of the clusters. The bluest objects in their sample showed strong correlation with the disk rotation of the galaxy, as traced by it's H\textsc{II} regions. As cluster color reddened, the rotation curve flattened, and their oldest sample of 14 clusters exhibited effectively no bulk rotation, indicating that the combined velocity dispersion of these clusters should be roughly 70 \kms (see their Figure 3). This affect was also identified in \citet{chandar+2002} (see their Figure 4), who find that the oldest clusters in their sample (defined as being older than $8\times10^8$ Gyr and a $|V_{\rm disk} - V_{\rm cluster}| \geq 70$ \kms) are best represented as having a significant halo component. They quote a $V/\sigma$ of 0.09 for the halo clusters, indicating that this population shows “marginal evidence for retrograde motion”. Note that the general conclusions of \citet{CS1991} and \citet{chandar+2002} are broadly consistent, despite poor agreement between the velocity measurements for individual clusters between the two studies.

\begin{table*}
\centering
\begin{tabular}{cccccccccccc}
\toprule
ID                 & \multicolumn{3}{c}{J22} & \multicolumn{3}{c}{SM07} & SR10 & \multicolumn{3}{c}{F14}         \\ 
                   & ID  & log(M) & log(Age) & ID   & log(Age) & log(M) & ID   & Age (Gyr) & {[}Fe/H{]} & log(M) \\ \midrule
AP01332868+3041349 & 271 & 4.76   & 8.39     & 95   &          &        & 1203 & 0.562     & -0.301     & 4.784  \\
AP01340286+3043207 & 603 & 3.01   & 6.83     & 281  & 9.21     & 5.56   & 1717 & 11.2      & -2.0       & 6.172  \\
AP01331924+3027503 &     &        &          &      &          &        & 1103 & 0.126     & -1.155     & 4.149  \\
AP01335212+3029036 &     &        &          &      &          &        & 1566 & 10        & -1.824     & 5.9    \\
AP01340197+3039377 & 718 & 2.14   & 6.82     & 273  & 9.8      & 6.11   & 1704 & 5.01      & -2.301     & 6.011  \\
AP01340801+3038380 & 649 & 3.49   & 7.99     & 316  & 10       & 6.27   & 1765 & 2.51      & -1.602     & 5.517  \\
AP01343022+3038127 &     &        &          &      &          &        & 2075 & 1.26      & 0.176      & 5.137  \\
AP01340021+3037468 & 709 &        &          & 261  & 8.96     & 5.48   & 1672 & 0.891     & 0.176      & 5.351  \\
AP01324221+3021141 &     &        &          &      &          &        & 714  & 11.2      & -1.602     & 5.793  \\
AP01332839+3036229 & 42  & 4.55   & 8.55     & 93   & 8.76     & 4.83   & 1199 & 0.178     & 0.477      & 4.123  \\
AP01333721+3034139 & 70  & 4.29   & 7.96     & 140  & 6.94     & 3.57   & 1311 & 0.158     & -1.602     & 4.356  \\
AP01335512+3047580 & 8   & 4.59   & 8.23     & 222  & 7.7      & 4.44   & 1601 & 0.631     & -1.456     & 4.955  \\
AP01341109+3047349 & 925 & 4.55   & 8.2      &      &          &        & 1822 & 5.62      & -1.602     & 5.686  \\
AP01341152+3034524 & 63  & 4.47   & 7.99     & 337  & 8.31     & 4.81   & 1829 & 1.78      & -2         & 5.377  \\
AP01332053+3049015 &     &        &          &      &          &        & 1115 & 3.55      & -2.301     & 5.634  \\
AP01332304+3050116 &     &        &          &      &          &        & 1138 & 1         & 0.176      & 4.984  \\
AP01333777+3057132 &     &        &          &      &          &        & 1322 & 0.794     & 0.176      & 4.745  \\
AP01342080+3057092 &     &        &          &      &          &        & 1965 & 1.12      & 0.176      & 5.068  \\
AP01342333+3101137 &     &        &          &      &          &        & 2001 & 7.94      & -2.301     & 5.759  \\
AP01343813+3029585 &     &        &          &      &          &        & 2145 & 0.891     & 0.176      & 5.024  \\ \bottomrule
\end{tabular}
\caption{We present measurements from the literature of our Gold sample clusters. We show the internal ID, log(M), and log(Age) from J22 \citep{Johnson2022} and SM07 \citep{Sarajedini2007}. We also show results from F14 \citep{Fan2014}, who use the ID system of SR10 \citep{SanRoman2010}. These measurements were done using photometry and SED fitting, and were optimized for younger ($\lesssim$1 Gyr), more solar metallicity clusters, which explains some of the inconsistencies between results.}
\label{tab:phot_crossmatch}
\end{table*}

\begin{figure*}
\centering
\includegraphics[width=0.8\textwidth]{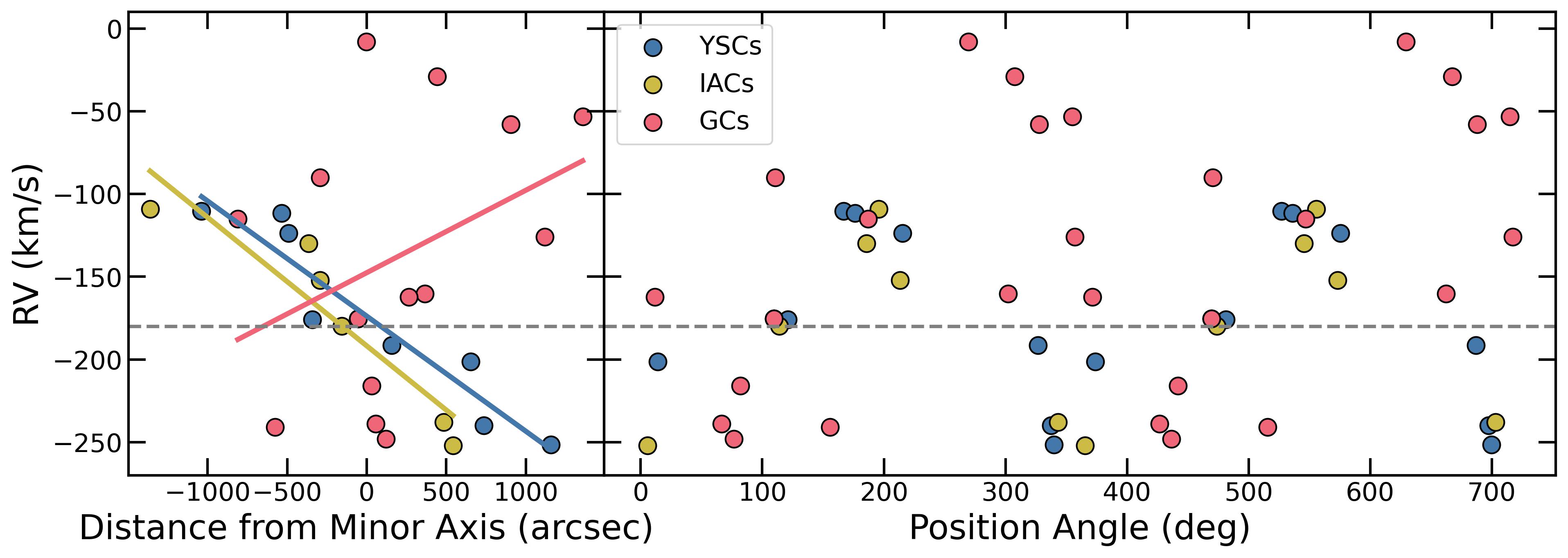}
\caption{Spatial kinematic assessment of our YSCs (blue), IACs (yellow), and GCs (yellow). In the left panel we plot each cluster's radial velocity vs its on-sky separation from M33's minor axis (113\degreespace east of north), where negative values are on the south side of the galaxy. We perform a linear fit to each population's relationship. In the right panel we plot each cluster's radial velocity vs its position angle relative to the disk major axis (23\degreespace east of north). IACs and YSCs have a coherent trend in both panels, while the GCs trends are highly scattered. The IACs and YSCs are therefore likely associated with the disk, while the GCs deviate substantially from this relation and are largey decoupled from the disk rotation. \textsc{Note:} we duplicated the clusters here to better show the cyclical structure of the right panel.}
\label{fig:pop_loc_sep}
\end{figure*}

In  the left panel of Figure~\ref{fig:pop_loc_sep} we plot this same trend, showing that the metal-rich, $\alpha$-poor IACs in our sample show clear kinematic consistency with the YSC's, which are both assumed to trace the rotation of the disk. However, this trend falls apart for the GCs, indicating that the metal-poor, $\alpha$-rich clusters are part of the same halo populations identified by \citet{CS1991} and \citet{chandar+2002}. 

\citet{beasley+2015} split spectroscopic sample into age bins and calculated metallicity gradients for the different populations. They find that all but their oldest clusters have essentially a flat gradient, with the oldest being distinctly negative, at least out to R$\sim$4.5 kpc, where it turns upwards again. We lack sufficient numbers of clusters across all to conclusively confirm this trend, but it does appear that our abundance gradients are roughly equivalent with flat for the low- and high-$\alpha$ populations.

\begin{figure}
\centering
\includegraphics[width=\columnwidth]{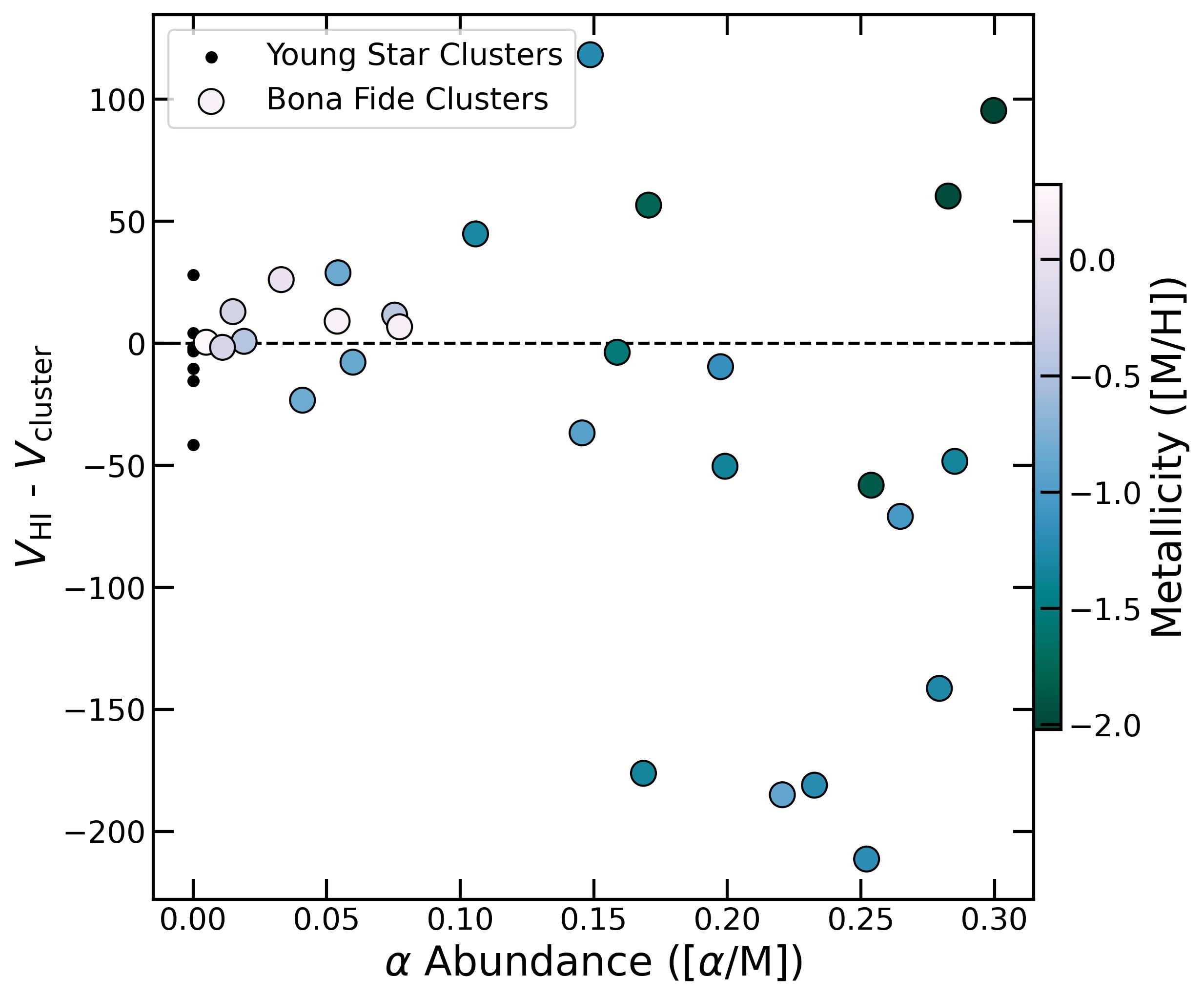}
\caption{Velocity of H\textsc{I} gas in M33 from \citet{Gratier2010} minus cluster velocity vs. cluster $\alpha$ abundance, colored by metallicity for all the clusters in our Bona Fide sample. We also show the YSCs as black dots, to show that they are generally consistent with their local disk kinematics.}
\label{fig:HI_comp}
\end{figure}

More interestingly, however, is the clear trend \citet{beasley+2015} between cluster metallicity and $|V_{\rm disk} - V_{\rm cluster}|$. Their Figure 18 shows that their oldest clusters (Age$\sim$10 Gyr) are essentially completely decoupled from the disk kinematics, with velocity dispersions ranging from roughly -200 to +100 \kms. We show the same plot for our Gold sample, though with $\alpha$ abundance on the x-axis, in Figure~\ref{fig:HI_comp}. This plot shows that the metal-poor, $\alpha$-rich clusters in our sample are, in general, kinematically dissociated with the disk, and suggests that the halo population in M33 is $\alpha$ enhanced.

\begin{figure*}
\centering
\includegraphics[width=0.8\textwidth]{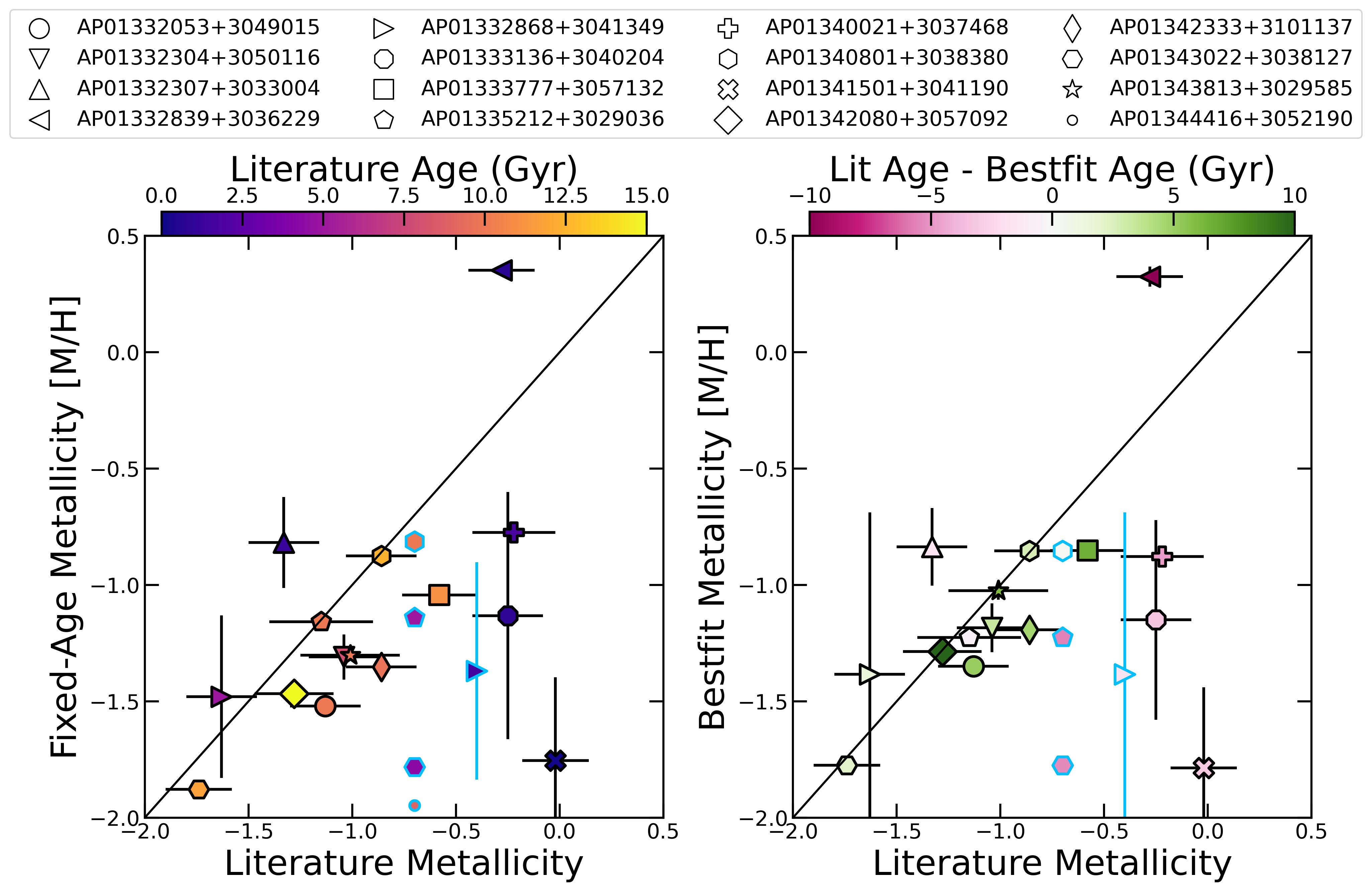}
\caption{Comparison of our results and those of \citet{Sarajedini2007} and \citet{beasley+2015}. On the left we plot the metallicity of our clusters when measured with the age fixed to the A-LIST grid value closest to the age measured by \citet{beasley+2015}, plotted versus the literature metallicity and colored by the literature age. On the right, we show the same sample, but now we plot the best-fit metallicity where age is allowed to vary, plotted again versus literature metallicity, and now colored by the difference between the literature age and our best fit age.}
\label{fig:lit_comparison}
\end{figure*}

Finally, in Figure~\ref{fig:lit_comparison}, we show direct comparison of our measured metallicities with those of \citet{beasley+2015} for clusters in our Bona Fide sample. The purpose of this plot is to show that when the age of the cluster is $\gtrsim$1 Gyr, our metallicity measurements are generally consistent with those of \citet{beasley+2015}, regardless of our measured age. This, along with the figures in Apppendix~\ref{app:ages}, show that our spectral analysis produces a relatively age-insensitive measurement of metallicity. Therefore, we conclude that our interpretations of the different sub-populations in the M33 cluster population is consistent with those of previous studies.

\subsection{Comparison to the MW's GC System} \label{sec:disk_comp}

\begin{figure}
\centering
\includegraphics[width=\columnwidth]{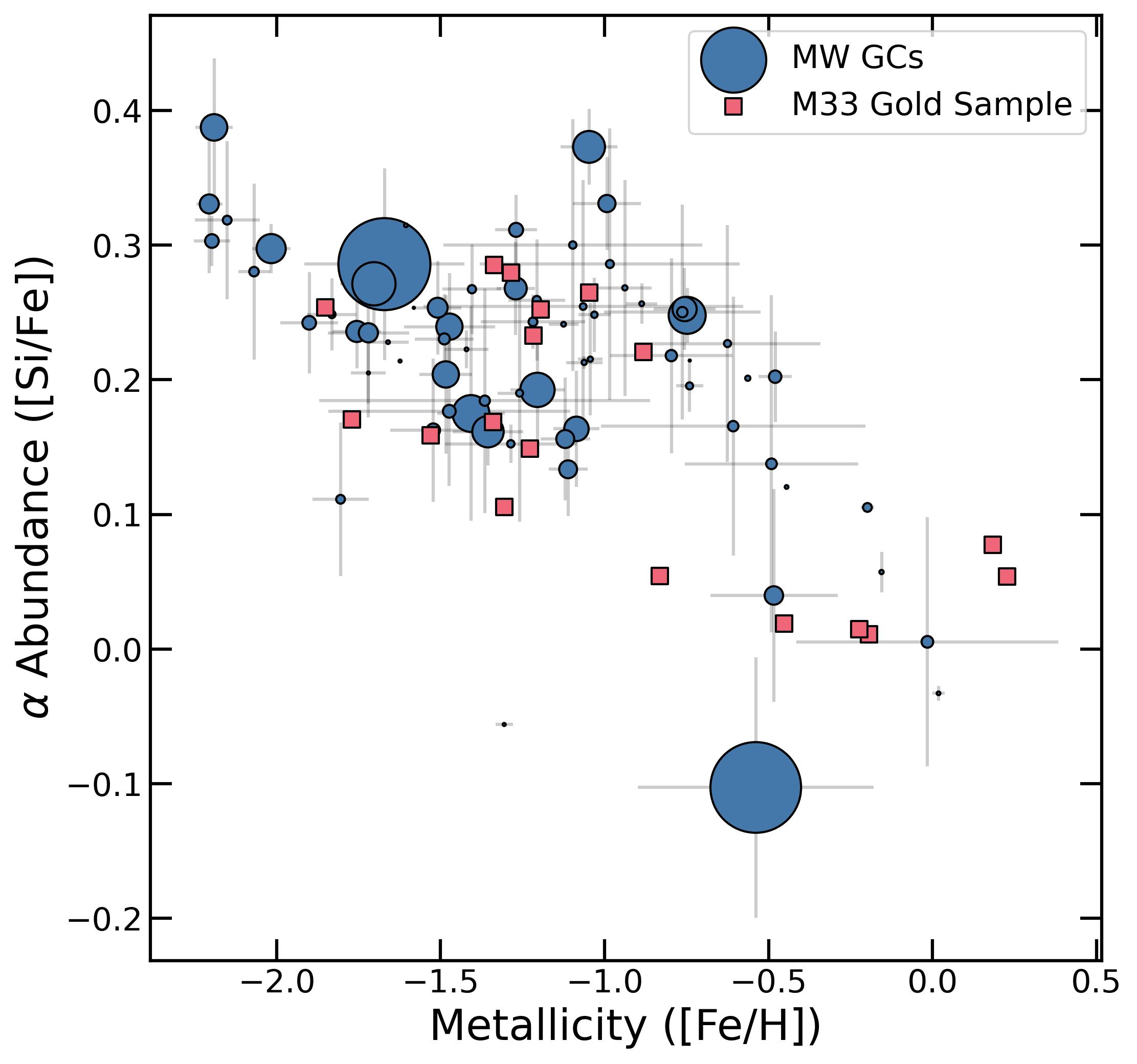}
\caption{Comparison of our results for the Gold Sample (red squares) with median abundances of individual stars in MW GCs from the APOGEE survey (blue circles -- the size is proportional to the number of cluster stars observed by APOGEE). We use [Si/Fe] as a proxy for $\alpha$ abundance as it is a relatively pure $\alpha$ element that is very precisely measured by APOGEE. This demonstrates that both galaxies show a spread in their $\alpha$ abundance at fixed metallicity, most notably around -1.2 dex. \textsc{Note:} the two MW GCs for which we have the most stars are $\rm \omega Cen$ (upper left) and M54 (lower right).}
\label{fig:MW_abundances}
\end{figure}

In Figure~\ref{fig:MW_abundances} we directly compare our M33 GC abundances to MW GC abundances as derived from the APOGEE survey. For this figure, we used the APOGEE targeting flags \citep{Beaton2021} to identify cluster members, then took the abundance of each cluster to be the median of the cluster stars. We estimated the intrinsic dispersion by maximising a Gaussian likelihood that includes each star’s abundance uncertainty. This accounts for measurement noise and describes the intrinsic star-to-star scatter, not the uncertainty on the median.

We use Silicon abundance ([Si/Fe]) as a proxy for total $\alpha$ abundance as it is among the purest $\alpha$-process elements, and is precisely measured by the APOGEE pipeline. Furthermore, Si is generally less affected by the abundance variations, particularly those in Mg and O, that are associated with multiple populations in GCs \citep[e.g.][]{Gratton2004, Bastian2018}. Since the relative numbers of stars belonging to a given generation in our APOGEE MW GC sample is not known, Si provides a better comparison for integrated light measurements that is less subject to systematic effects than other prominent $\alpha$ elements. Our comparison is intended to place the M33 measurements in the context of the observed abundance patterns of Galactic GCs, rather than to establish a strict one-to-one correspondence between stellar and integrated-light abundance measurements. Such a comprehensive comparison is beyond the scope of this work.

We can see from this figure that there is a spread in abundance for both MW and M33 clusters between -1.5 and -1.0 dex in metallicity, with a potential bimodality in the population for both galaxies. Such a spread in $\alpha$ abundances for MW GCs has been argued as a sign of a population accreted from presumably less massive systems \citep[e.g.,][]{Horta2020}. Moreover, GCs residing in the MW halo have been dynamically associated with the Sagittarius Dwarf Spheroidal galaxy \citep[e.g.,][]{Law&Majewski2010,Bellazzini2020} and the \emph{Gaia}-Sauasage/Enceladus galaxy \citep[e.g.,][]{Massari2019}. While we speculate in Sec\ref{sec:retrograde} that some of these M33 clusters have anomalous kinematics, suggestive of being accreted, the spread in $\alpha$ abundances observed in the M33 GCs implies, at the very least, that the M33 GCs formed in a range of star formation conditions. 

Regardless of the potential origin of this spread, the fact that M33 also hosts a population of $\alpha$-enhanced GCs that are chemically akin to MW GCs suggests that both galaxies formed stars (or at least, GCs) in similar ways at early times, even though M33 is an order of magnitude lower mass than the MW. This either means that galaxies formed GCs in the early Universe in similar ways, regardless of the mass of the system, or that the MW and M33 were much more alike at early times, and it was their evolution across cosmic time that differed. As measurements of proper motion continue to suggest that M33 is experiencing its first infall into M31 and the Local Group \citep[e.g.,][]{vanderMarel2019}, the latter scenario seems more likely.

Previous studies of the metallicity gradient of M33 have revealed that the young stars exhibit a negative radial gradient \citep[e.g.,][]{Monteverde1997}, or even a two-slope negative gradient that flattens at larger radii \citep[e.g.,][]{Magrini2007}. A two-slope gradient is similar to what is observed for the young clusters in the MW \citep[e.g.,][]{Donor2020}. Unfortunately, we do not have enough IACs to conclusively measure a metallicity gradient for our young clusters.

The GCs in our Gold sample, however, are much more similar to the GCs found in the MW in two major ways. First of all, as shown in Figure \ref{fig:HI_comp}, the more metal-poor GCs in M33 have kinematics inconsistent with belonging to the disk. This is similar to the MW GCs, suggesting that these structures were not formed as part of the disk, either forming before a disk was established or accreted from a different system via a merger.

\subsection{Potentially Retrograde Clusters} \label{sec:retrograde}

With only projected sky positions and line-of-sight velocities available, the three-dimensional spatial distribution and kinematics of M33's star cluster system cannot be determined directly. However, by comparing the radial velocity of the clusters to the rotation of M33's disk, we can gain insight on the probable orbital sense of the clusters, and find which clusters have the largest departures from the kinematics of the disk.

We first assume that the clusters lie in the disk of M33, allowing their projected positions and velocities to be deprojected assuming a geometry for M33 and its rotation curve. We test several different geometries. The implications and limitations of this assumption are discussed in Section~3.5 of \citet{Anguiano2025}.

Assuming that the clusters follow circular orbits in a thin disc, the observed line-of-sight velocity at a given position can be written as

\begin{equation}
    V_{\mathrm{los}} = V_{\mathrm{sys}} + V_{\mathrm{rot}}(R)\,\sin i\,\cos\theta,
    \label{eq:projected_rotation}
\end{equation}

where \(V_{\mathrm{los}}\) is the measured line-of-sight velocity of the cluster, \(V_{\mathrm{sys}}\) is the systemic velocity of M33, and \(V_{\mathrm{rot}}(R)\) is the rotational velocity in the disk at the deprojected galactocentric radius \(R\). We adopt $V_{\rm sys} = -179$ \kmspace following \citet{Koch2018}. The inclination \(i\) is defined as the angle between the line of sight and the normal to the disk, while \(\theta\) is the azimuthal angle in the plane of the disc, measured from the projected major axis. 

Following \citet{Rubin_Ford1970}, the geometric projection factor is given by
\begin{equation}
    \cos\theta = \frac{X}{R},
    \label{eq:costheta}
\end{equation}

where \(X\) is the coordinate along the projected major axis. The deprojected galactocentric radius is calculated as:

\begin{equation}
    R = \left[X^{2}+\left(\frac{Y}{\cos i}\right)^{2}\right]^{1/2},
    \label{eq:deprojected_radius}
\end{equation}

where \(Y\) is the coordinate along the projected minor axis. 

\begin{figure}
\centering
\includegraphics[width=\columnwidth]{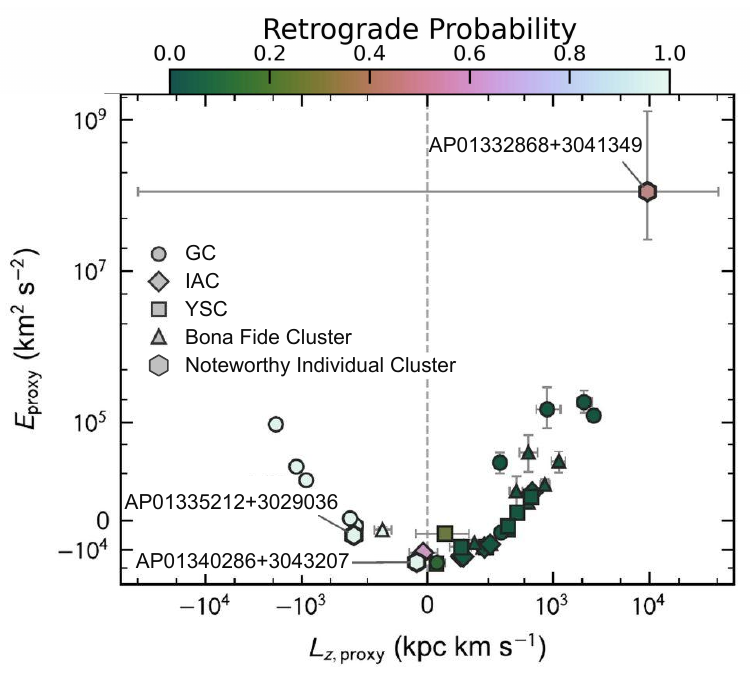}
\caption{The probability of our Bona Fide sample clusters orbiting retrograde, based on radial velocity projection method outlined in \citet{Anguiano2025}. A positive value on the x-axis is an indicator of prograde motion, while negative values are more likely to be retrograde. All of the IACs (diamonds) and YSCs (squares) are strongly prograde, while all the strongly retrograde clusters are GCs (circles). AP01335212+3029036, AP01340286+3043207, and AP01332868+3041349 (hexagons), the three clusters discussed in Section~\ref{sec:oddballs}, are listed from most negative to most positive. The large error bars on AP01332868+3041349 are due to its proximity to the minor axis. \textsc{Note:} the scaling on both axes is asinh to accommodate the wide range of results.}
\label{fig:retrograde_probs}
\end{figure}

We test several geometries for M33 and its kinematics, following the major-axis position angle, inclination, and rotation curve measurements from \citep{Ciardullo2004, Kam2015, Kam2017, Koch2018, Cullinane2023}. We assume a distance to M33 of \(840~\mathrm{kpc}\) when converting angular separations into physical galactocentric radii. We present the results of this calculation averaged across the different geometries in Figure~\ref{fig:retrograde_probs}.

According to \citet{Anguiano2025}, two of our clusters | AP01332868+3041349, and AP01335212+3029036 | have a high probability of retrograde motion; both of these are included in our discussion in Section~\ref{sec:oddballs}. The first of these has the largest $-L_z$ and $E$ in their sample, though they used the value of -192 \kmspace from \citet{beasley+2015} for their calculations. This calculation using our measurement produces an even larger energy, further enhancing the argument that this object has an accreted origin, despite having a chemical abundance in line with M33's other disk-like GCs. \citet{Anguiano2025} further argue that this cluster likely is not bound to M33, and could be a dwarf spheroidal within the Local Group. The second cluster, AP01335212+3029036, has the lowest kinetic energy of any of the halo-like clusters (squares) in Figure~\ref{fig:retrograde_probs}, but it's location on the disk gives it a high retrograde probability. 

Other clusters in this high retrograde probability region include \citet{beasley+2015} GC candidates such as AP01332053+3049015, AP01332304+3050116, AP01333777+3057132, AP01342080+3057092, AP01342333+3101137, as well as AP01340286+3043207, which is one of the notable individual clusters discussed in Section~\ref{sec:oddballs}.

We caution that these retrograde probabilities neither directly measure the clusters’ orbital senses nor robustly distinguish disk from halo membership. The clusters with the highest retrograde probabilities suggest departures from disk kinematics, but determining their orbital senses requires 3D velocity measurements. Moreover, retrograde motion does not uniquely imply accretion into M33, as \textit{in situ} formation followed by dynamical heating remains a viable alternative. Nevertheless, this collection of clusters provides our strongest evidence for accretion into the M33 halo, combining unusual abundances and kinematics with projected radial velocities that lack coherent motion relative to M33’s \textit{in situ} stellar population. Star-by-star SED fitting with \textit{JWST} and proper-motion measurements with \textit{HST} could help establish their origins, while broader spectroscopic studies of M33’s cluster population will provide the context needed to interpret these results.

\subsection{Noteworthy Individual Clusters} \label{sec:oddballs}

\begin{figure*}
\centering
\includegraphics[width=0.8\textwidth]{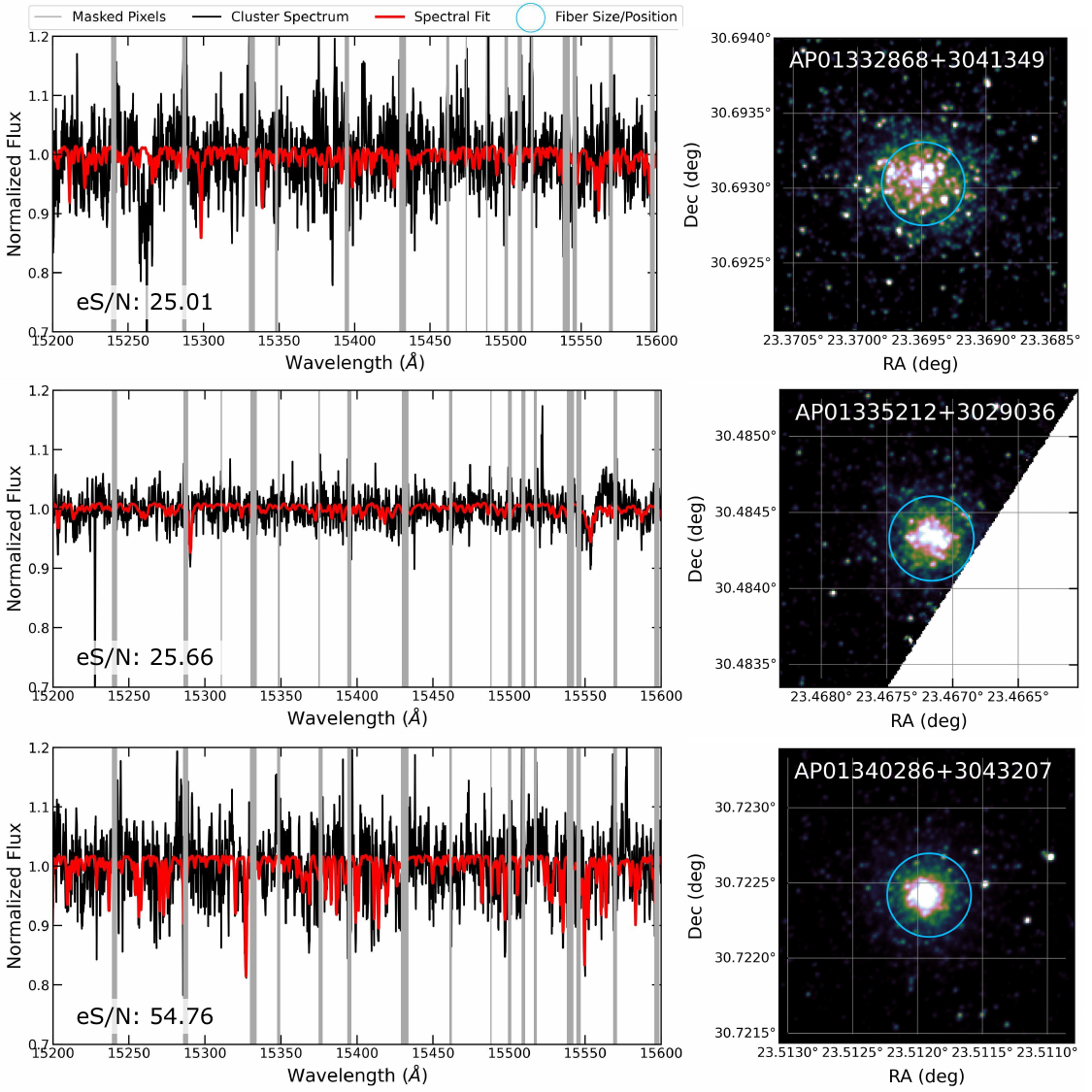}
\caption{Comparison of the spectral fit and imagery of the noteworthy clusters discussed in Section~\ref{sec:oddballs}, three kinematically hot GCs that are the strongest candidates for accretion into M33. This figure is formatted the same as Figure~\ref{fig:photspeccomp}. All imagery is from the PHATTER survey \citep{PHATTER}.}
\label{fig:oddballs}
\end{figure*}

While the majority of the clusters in our sample are naturally grouped into the three populations described in Section~\ref{sec:populations}, three objects warrant individual discussion because their unusual chemistry and kinematics indicate a possible accreted origin. These clusters have also been studied extensively over the past several decades under a variety of catalog designations, making it difficult to connect results across the literature. For convenience, we list the most commonly used identifiers for each object and provide a link to its SIMBAD entry. We present \textit{HST} images of the clusters where possible from the PHATTER survey \citep{PHATTER} in Figure~\ref{fig:oddballs}.

In general, we find that our cluster parameters are most consistent with previous integrated-light spectroscopic measurements and resolved stellar population analyses based on color-magnitude diagrams (see Figure~\ref{fig:lit_comparison} for comparison with \citet{beasley+2015}). In contrast, ages derived from integrated-light SED fitting are often more discrepant, likely due to the strong age-metallicity degeneracy inherent to these models and the assumptions adopted in their fitting procedures. Several of these studies were designed primarily to characterize the young, metal-rich cluster population of M33, and therefore their adopted metallicity grids are not always well matched to old, metal-poor GCs. As a result, some SED-based age estimates likely underestimate the true ages of the clusters. We nevertheless include the multi-band SED fitting results of Ma et al., which provide an additional independent integrated-light constraint and generally offer useful comparisons to our measurements.

\subsubsection{AP01332868+3041349}
\textbf{Common identifiers:} U77, CBF90, SM95, SR1203.\footnote{\url{https://simbad.cds.unistra.fr/simbad/sim-id?Ident=\%5bCS82\%5d\%20U77&submit=submit\%20id}}

This cluster has chemistry typical of the Population 1 GCs ($\mathrm{[M/H]}=-1.34\pm0.09$, $\mathrm{[\alpha/M]}=0.17\pm0.01$), but stands out dramatically in its kinematics. Despite lying only $\sim2$ kpc from the center of M33, we measure a radial velocity of $-8.0\pm0.5$~\kms, the most redshifted cluster in our Gold sample and nearly at rest with respect to the Milky Way. Morphologically, the cluster is somewhat more diffuse than most GCs in our sample, with numerous resolved stars extending beyond the 2\arcsec\space APOGEE fiber (left panel of Figure~\ref{fig:oddballs}).

Previous studies consistently classify this object as an old, metal-poor GC. Published spectroscopic and CMD-based metallicities span approximately $-1.8$ to $-0.8$ dex \citep{Brodie1991, Sarajedini1998, Ma2004, chandar2006, beasley+2015, Larsen2022}, with ages that are typically older than $\sim5$ Gyr \citep{CS1983, CS1988, beasley+2015}, though \citet{chandar+2002} and \citet{Ma2002a} find an age of just $\sim1.5$ Gyr. Furthermore, \citet{Larsen2022} measure $\rm [Mg/Fe]=0.147\pm0.093$, which is consistent with our $\alpha$ abundance measurement.

The largest discrepancy in the literature is the radial velocity, with previous measurements of $-253$, $-216$, and $-192$~\kmspace from \citet{CS1991}, \citet{chandar+2002}, and \citet{beasley+2015}, respectively, which differ greatly from our measurement. To investigate these radial velocities, we repeated our spectral fits while fixing the radial velocity to the published values. Both solutions produced significantly poorer fits and required implausibly large internal velocity dispersions ($\sigma_{\rm int}=39$--$76$~\kms), whereas our preferred solution yields $\sigma_{\rm int}=1.7$~\kms. We therefore retain our measured radial velocity. 

The combination of its unusual motion and otherwise typical GC chemistry makes this object one of the strongest candidates for an accreted halo cluster in our sample. 

\subsubsection{AP01340286+3043207}
\textbf{Common identifiers:} MKK33, SM281, SR1717.\footnote{\url{https://simbad.cds.unistra.fr/simbad/sim-basic?Ident=\%5BMKK98\%5D+M+33+33&submit=SIMBAD+search}}

This cluster exhibits the most extreme chemistry in our Bona Fide sample, with $\mathrm{[M/H]}=-1.85\pm0.05$ and $\mathrm{[\alpha/M]}=0.25\pm0.03$. It is both the most metal-poor cluster in the Gold sample and among the most $\alpha$-enhanced. Although located only 1.2 kpc from the center of M33, its radial velocity ($-162.2\pm0.5$~\kms) is significantly offset from the local H\textsc{I} rotation, and it has the largest measured internal velocity dispersion in the Gold sample ($9.7\pm0.9$~\kms). HST imaging reveals a compact, centrally concentrated morphology characteristic of a classical GC.

Previous work has consistently identified this object as a GC, with the only previous integrated-light spectroscopy coming from \citet{sharina+2010}. They measured an age of $1.2\pm0.7$ Gyr, though with super-solar metallicity $\mathrm{[Z/H]}=0.2\pm0.2$ and high $\alpha$ abundance $\mathrm{[\alpha/Fe]}=0.3\pm0.3$. \citet{Ma2004} measure an age of 1.61 Gyr and metallicity $\mathrm{[Fe/H]}=-1.51\pm0.12$.

The only radial velocity measurement in the literature is $-193\pm16$~\kms from \citet{sharina+2010}. Their measurement is close to the local H\textsc{I} velocity, and may therefore be affected by contamination from the surrounding disk light. However, this value differs from our measurement by only $\sim2\sigma$ and is therefore broadly consistent.

Taken together, the cluster's unusually metal-poor, $\alpha$-enhanced chemistry and its kinematic offset from the rotating disk distinguish it from the rest of the Population 1 GCs. It is therefore another strong candidate for an accreted member of the M33 halo.

\subsubsection{AP01335212+3029036}
\textbf{Common identifiers:} H38, MKK22, CBF104, SM206, SR1566.\footnote{\url{https://simbad.cds.unistra.fr/simbad/sim-basic?Ident=\%5BSSA2010\%5D+1566&submit=SIMBAD+search}}

Chemically, this cluster is representative of the intermediate-$\alpha$ Population 1 GCs ($\mathrm{[M/H]}=-1.23\pm0.04$, $\mathrm{[\alpha/M]}=0.15\pm0.05$). Its distinguishing feature is its kinematics: we measure a radial velocity of $-240.8\pm0.3$~\kms, making it one of the most blueshifted GCs in the sample relative to the local disk. Figure~\ref{fig:oddballs} shows the cluster's compact morphology in PHATTER imaging.

This object has been extensively studied both spectroscopically and photometrically. Published metallicities generally fall between $-1.5$ and $-1.15$ dex \citep{Cohen1984, Schommer1991, Sarajedini1998, beasley+2015, Larsen2022}. Ages from spectra and CMDs range from $>$1 to 10 Gyr \citep{Schommer1991, chandar+2002, Park2009, beasley+2015}. \citet{Ma2004} use 13-band SED fitting to measure a metallicity $\rm [Fe/H]=-0.84\pm0.036$ and an age of $5.0\pm0.7$ Gyr.

\citet{Larsen2022} measure $\rm [Mg/Fe]=+0.144\pm0.136$, which is right in line with our $\alpha$ abundance.

Reported radial velocities range from approximately $-255$ to $-195$~\kmspace \citep{Cohen1984, Schommer1991, chandar+2002, Larsen2002, beasley+2015}, in reasonable agreement with our value.

Overall, this object appears to be a typical old GC whose primary distinguishing characteristic is its unusually large blueshift relative to the M33 disk. Unlike the previous two clusters, its chemistry is otherwise unremarkable within the Population 1 sample. Nevertheless, it too is a candidate for accretion into M33's stellar halo.

\section{Summary and Conclusion} \label{sec:conclusion}

We present an analysis of 93 APOGEE spectra of cluster candidates in M33, representing the first high-resolution, near-infrared integrated-light spectroscopic study of the M33 cluster system. Through an \textit{HST} archival image search and cross matching with literature catalogs, we narrow the sample to 37 bona fide clusters, shown in Table~\ref{tab:catalogs}. Utilizing a slightly modified visit-stacking routine from \citetalias{Gibson2025}, we perform full-spectrum fitting using pPXF and A-LIST models. We define the 20 clusters with eS/N $\gtrsim15$ as our Gold sample and present their kinematics, metallicities ([M/H]), and $\alpha$-abundances ([$\alpha$/M]) in Table~\ref{tab:results}. Our analysis naturally separates the Bona Fide cluster sample into three populations based on their kinematics, chemical abundances, and approximate ages inferred from visual inspection of the spectra:

\begin{itemize}
    \item \textbf{YSCs} | Young Star Clusters: we estimate that these are likely $\lesssim$1 Gyr old, as their spectra are dominated by broad Brackett absorption features from bright A-type stars.
    \item \textbf{IACs} | Intermediate-Age Clusters: we estimate these are likely $\sim$1--3 Gyr old, as their spectra are rich in narrow metal absorption lines dominated by F- and G-type stars. These have approximately solar metallicities and $\alpha$ abundances.
    \item \textbf{GCs} | Disk- and Halo-like Globular Clusters: we estimate these are likely old ($\gtrsim$5 Gyr), given the lack of prominent spectral features. They are metal-poor ([M/H] $\lesssim$ -1.0), and span a broad range of chemical abundances and kinematic properties.
\end{itemize}

Our primary conclusions are as follows:

\begin{itemize}
    \item The YSC and IAC populations closely follow the rotation of the M33 disk, as traced by the H\textsc{I} velocity field of \citet{Gratier2010} (see Figure~\ref{fig:HI_comp}).
    
    \item The GC population spans metallicities of $\rm -1.8 \lesssim [M/H] \lesssim -0.6$, and $\rm 0.05 \lesssim [\alpha/M] \lesssim 0.3$, with a spread in $\rm [\alpha/M]$ of $\sim$0.2 dex at $\rm[M/H]\approx-1.2$. This is similar to the trends seen in MW GCs, and implies a similar set of formation scenarios for the two galaxies clusters.
    
    \item The most $\alpha$-enhanced GCs exhibit kinematics inconsistent with disk rotation and are likely members of the dispersion-supported stellar halo of M33. Combined with their chemical properties, these results suggest that the halo GC population has a mixed \textit{in-situ} and accreted origin.
    
    \item Our radial velocity and abundance measurements, which are effectively age-insensitive for clusters older than $\sim$1 Gyr (see Appendix~\ref{app:ages}), broadly agree with previous spectroscopic studies.
    
    \item Three clusters exhibit unusual kinematics, chemistry, or spatial locations relative to the broader GC population and are discussed individually in Section~\ref{sec:oddballs}.
    
    \item Applying the radial velocity projection methodology of \citet{Anguiano2025}, we find strong evidence that several halo-like GCs | including many of the GC candidates identified by \citet{beasley+2015} | are potentially on retrograde orbits, providing further evidence that they may belong to an accreted halo subsystem.
\end{itemize}

Taken together, these results demonstrate that the M33 cluster system is not a homogeneous population, but instead contains both disk and halo components with distinct chemical and kinematic signatures. The presence of chemically and kinematically distinct halo GCs suggests that the assembly history of M33 has been more complex than expected for a galaxy of its mass -- either because of mergers or complex early star formation scenarios. Future high-resolution imaging, expanded integrated-light spectroscopic surveys, and more comprehensive stellar population models will allow the origins of these clusters to be tested in greater detail and provide new constraints on the formation and chemical evolution of M33.

\begin{acknowledgments}

We appreciate comments and suggestions from the anonymous referee that helped improve the manuscript. 

Funding for the Sloan Digital Sky Survey IV has been provided by the Alfred P. Sloan Foundation, the U.S. Department of Energy Office of Science, and the Participating Institutions. 

SDSS-IV acknowledges support and resources from the Center for High Performance Computing  at the University of Utah. The SDSS website is www.sdss4.org.

SDSS-IV is managed by the Astrophysical Research Consortium for the Participating Institutions of the SDSS Collaboration including the Brazilian Participation Group, the Carnegie Institution for Science, Carnegie Mellon University, Center for Astrophysics | Harvard \& Smithsonian, the Chilean Participation Group, the French Participation Group, Instituto de Astrof\'isica de Canarias, The Johns Hopkins University, Kavli Institute for the Physics and Mathematics of the Universe (IPMU) / University of Tokyo, the Korean Participation Group, Lawrence Berkeley National Laboratory, Leibniz Institut f\"ur Astrophysik Potsdam (AIP),  Max-Planck-Institut f\"ur Astronomie (MPIA Heidelberg), Max-Planck-Institut f\"ur Astrophysik (MPA Garching), Max-Planck-Institut f\"ur Extraterrestrische Physik (MPE), National Astronomical Observatories of China, New Mexico State University, New York University, University of Notre Dame, Observat\'ario Nacional / MCTI, The Ohio State University, Pennsylvania State University, Shanghai Astronomical Observatory, United Kingdom Participation Group, Universidad Nacional Aut\'onoma de M\'exico, University of Arizona, University of Colorado Boulder, University of Oxford, University of Portsmouth, University of Utah, University of Virginia, University of Washington, University of Wisconsin, Vanderbilt University, and Yale University.

\end{acknowledgments}

\facilities{Sloan (Infrared), HST (ACS)}

\software{Astropy \citep{2013A&A...558A..33A}, \textsc{ppxf} \citep{pPXF_1, pPXF_2, pPXF_3}}

\appendix

\section{How well can we derive ages?} \label{app:ages}

To strengthen our argument that ages are poorly constrained by fitting to A-LIST templates, we have performed two tests.

\begin{figure*}
\centering
\includegraphics[width=0.8\textwidth]{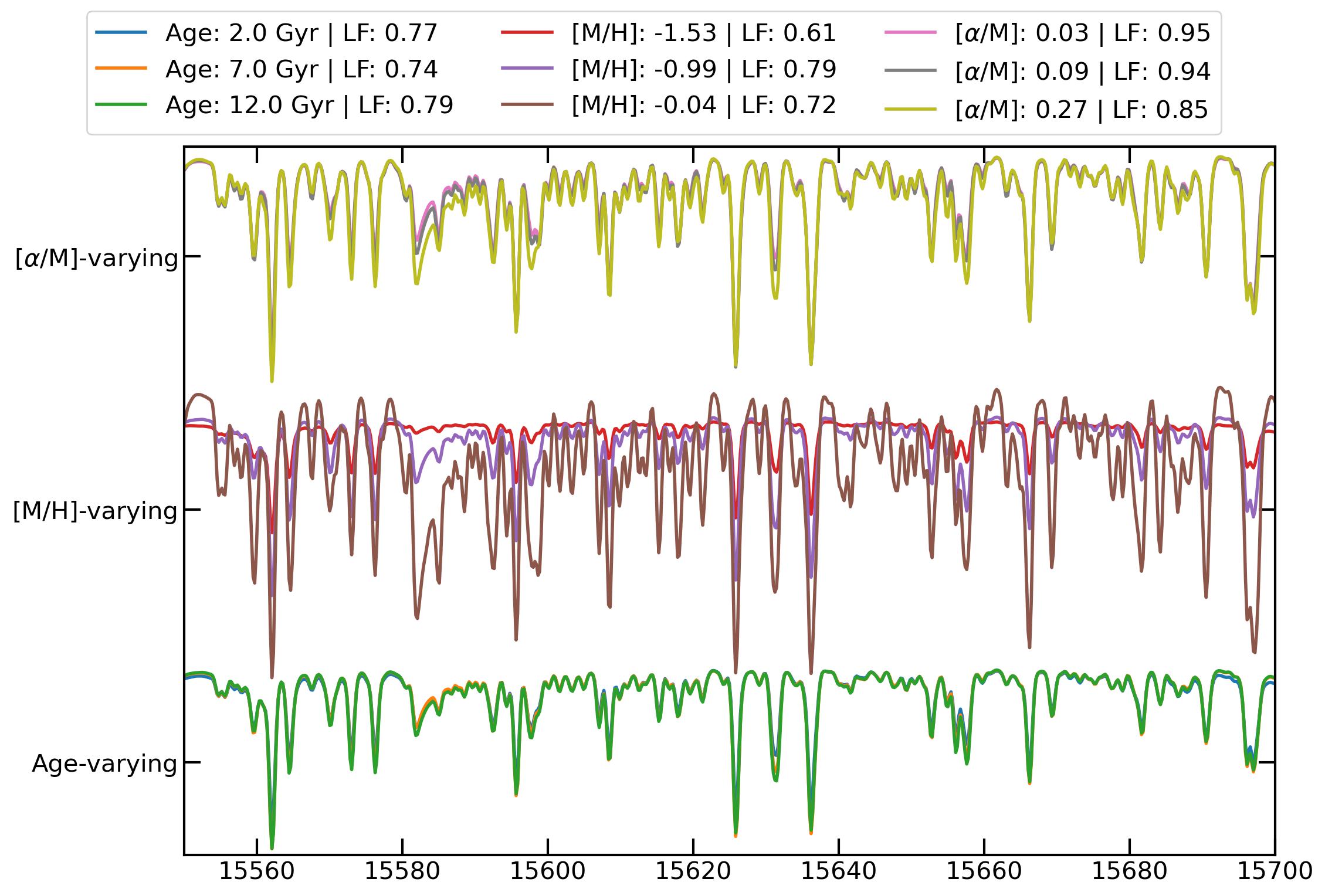}
\caption{We show nine different A-LIST templates plotted here, broadened by a gaussian velocity dispersion of 2 \kms. The bottom row shows three templates that share a metallicity of [M/H]=-1.0 and $\alpha$-abundance [$\alpha$/M]=0.2, but their ages vary. There is almost no difference at any point between these templates. The middle row shows three templates that share Age=12 Gyr, [$\alpha$/M]=0.2, with varying [M/H]. These templates have clear distinctions, with an obvious trend from low- to high-metallicity. Finally, the top row shows three templates with Age=12 Gyr, [M/H]=-0.6, and varying [$\alpha$/M]. These vary decently, especially in strong $\alpha$ features like the one at 15630 angstrom.}
\label{fig:template_ages}
\end{figure*}

The first test is shown in Figure~\ref{fig:template_ages}, where we display the variation between templates with respect to age, metallicity, and $\alpha$ abundance. The bottom row shows three templates that vary only in Age, and are nearly identical despite being 2.0, 7.0, and 12.0 Gyr old. The variations in metallicity and $\alpha$ abundance are far greater, showing our inability to measure ages confidently.

For the second test, we fit a set of mock observations to test our ability to recover metallicity and $\alpha$ abundance with respect to the age, velocity dispersion ($\sigma_{int}$) and eS/N. We generated 207 mock observations in a grid of values: $\sigma_{int}$ = 2, 5, \& 10 km/s; eS/N = 15, 30, \& 50; Age = 2, 7, \& 12 Gyr; [M/H] = -1.9, -1.0, \& 0.0; and [$\alpha$/M] = 0.1, 0.2, \& 0.3. Note that we only use A-LIST templates in this grid with a combination of Age, [M/H], and [$\alpha$/M] that have a luminosity fraction greater than 0.35 (see Section~\ref{sec:models}). The mock observations were created by taking a template, shifting its velocity to -180 \kms, and broadening it by the $\sigma_{int}$ value. To then make the mocks realistic, we use the noise, sky, and mask arrays from real Gold-sample spectra. We add in the noise and sky flux scaled to produce the desired eS/N (similar to the process outlined in Section~3.5.1 of \citetalias{Gibson2025}), and rerun the masking routine on the observation to produce very realistic mock observations across the desired parameter space.

We performed four fits to each mock observation: once where we use all A-LIST templates (as defined in Section 4), and three times where we only include A-LIST templates with ages of 2, 7, and 12 Gyr, respectively. This way, we could assess the impact of a) allowing pPXF to use whichever age templates it needs to produce the best fit, b) only allowing pPXF to use the templates with the actual input age of the mock observation, and c) intentionally getting the age incorrect.

\begin{figure*}
\centering
\includegraphics[width=0.49\textwidth]{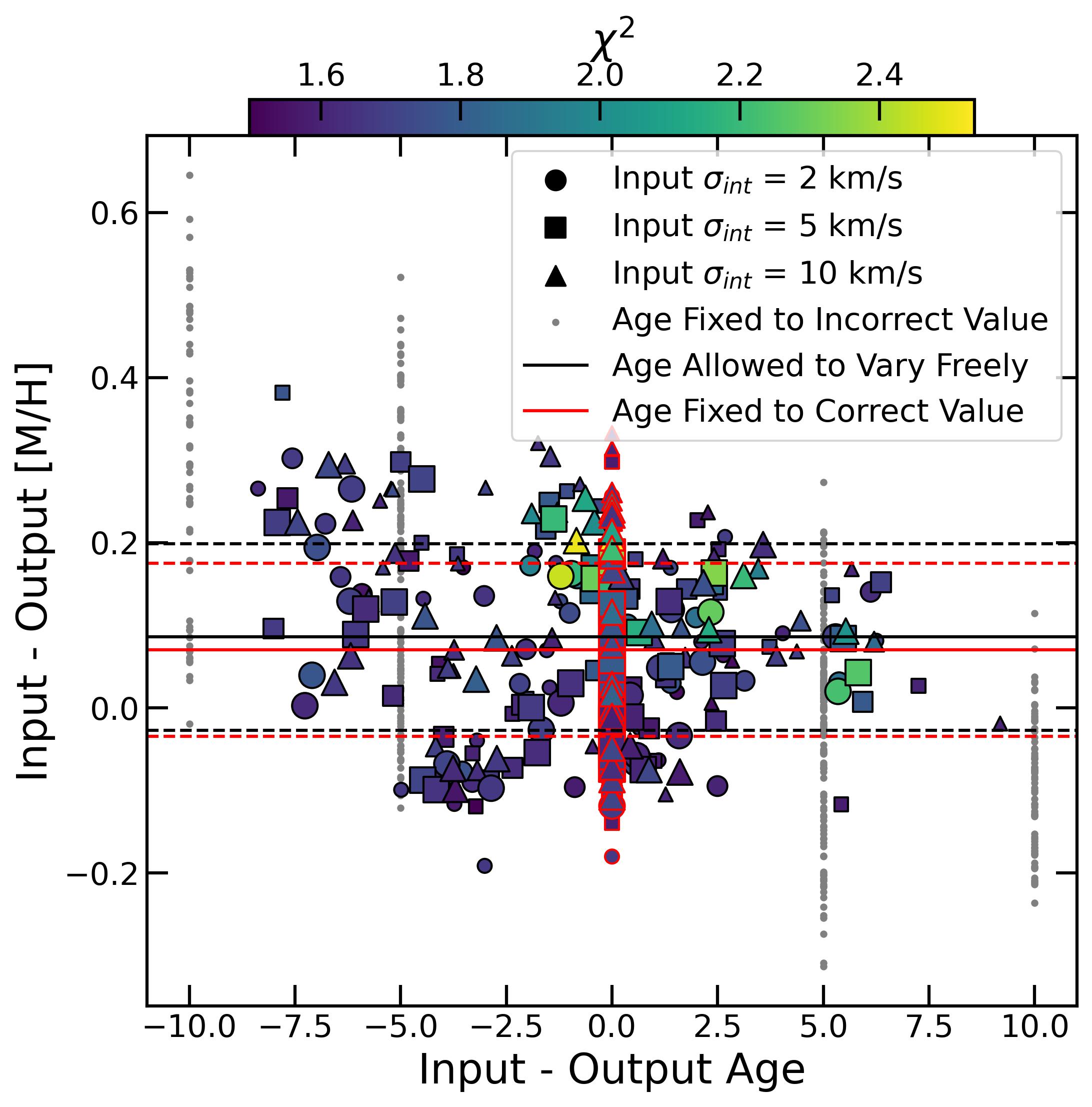}
\includegraphics[width=0.49\textwidth]{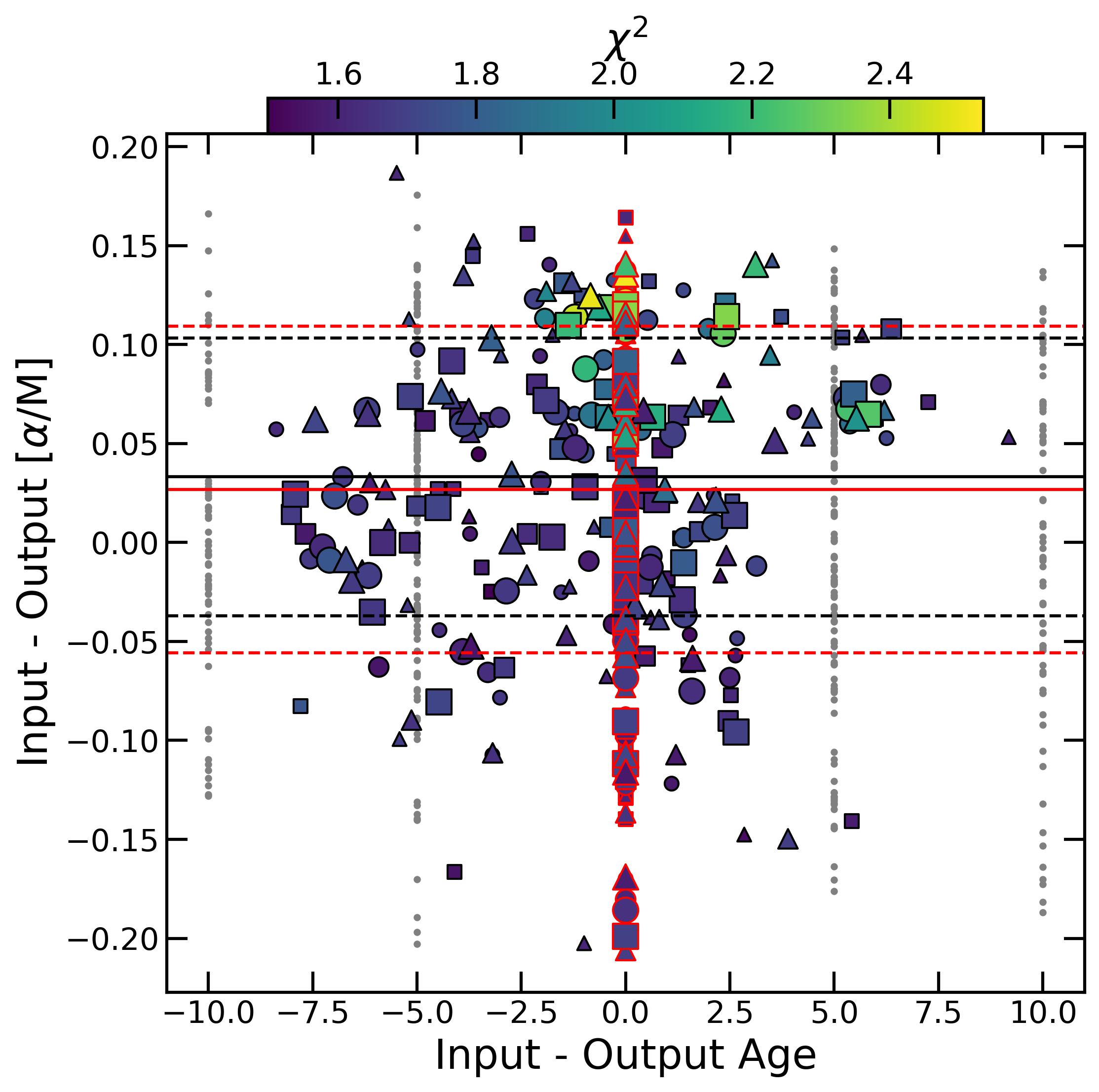}
\caption{Results of our mock observation abundance recovery tests. Marker shape indicates the input velocity dispersion, and marker size indicates the eS/N (larger sizes have higher eS/N). Markers with black outlines are results where we allow pPXF to measure the age; markers with red outlines are results where we fix age to the correct input value. Markers are colored by the reduced $\chi^2$ of the fit. We show the bias and standard deviation on the abundance recovery as solid and dashed lines for the varying- and correct-age cases in black and red, respectively. Results from the case where we intentionally fix the age to an incorrect value are shown as small gray dots. This test shows that our ability to accurately recover the abundances of our spectra is the same in the case where we know the age exactly and in the case where we allow pPXF to determine the best fit age, and significantly better than in the cases where we fix the age to an incorrect value.}
\label{fig:fixed_ages}
\end{figure*}

The results of this test are shown in Figure~\ref{fig:fixed_ages}. This test demonstrates that our ability to recover the input abundance is similar in the case where we allow pPXF to measure the best fit age (bias and standard deviation of 0.086±0.113 [M/H] and 0.033±0.070 for [$\alpha$/M]), and in the case where we fix the age to the correct value (bias and standard deviation of 0.070±0.105 [M/H] and 0.026±0.082 for [$\alpha$/M]). Additionally, our abundance recovery is most accurate in the case where eS/N is greater than 30, $\sigma_{int}$ is less than 5 \kms, and Age is 7 Gyr | the most likely scenario for our Gold sample GCs | with a bias and standard deviation of roughly 0.05±0.1 for metallicity. Our $\alpha$ abundance recovery had a similar accuracy, around 0.03±0.075, across the three ages we tested.

Furthermore, it is evident that fixing the age to an incorrect value produces systematically poorer abundance recovery than in either the varying- or correct-age cases. Therefore, we assert that our abundance measurements are effectively age insensitive, in the sense that when the cluster age is between 2 and 12 Gyr, we produce accurate metallicity and $\alpha$ abundance measurements regardless of the Age that pPXF measures, and demonstrate that allowing pPXF to determine the age universally produces more accurate abundances than fixing the age to an incorrect value.



\section{Cluster Images from CFHT} \label{app:pandas}

To visually identify cluster candidates, we largely used existing \textit{HST} imagery. However, such data is largely confined to M33's star forming disk. The most complete imagery to date of the halo of M33 come from the Pan-Andromeda Archaeological Survey, or PAndAS, a g- and i-band CFHT/MegaCam imaging survey \cite{McConnachie2010}. We use the i-band images to visually inspect clusters in the halo. The images were obtained through the Canadian Astronomy Data Centre (CADC) archive and processed using the MegaPipe pipeline \citep{MegaPipe}. 

\begin{figure*}
\centering
\includegraphics[width=0.8\textwidth]{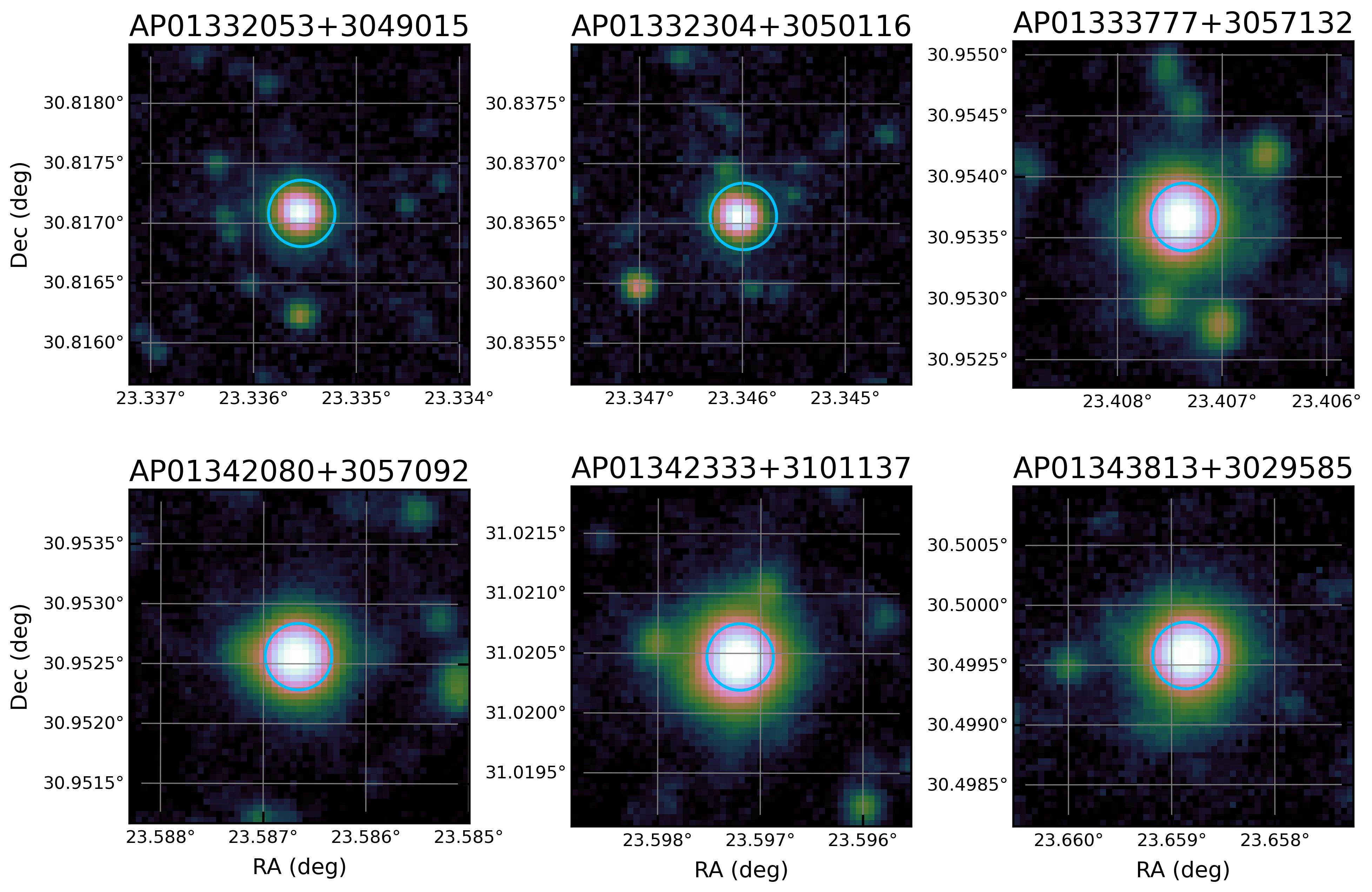}
\caption{CFHT MegaCam imagery \citep[MegaPipe,][]{MegaPipe} from the Pan-Andromeda Archaeological Survey \citep[PAndAS][]{McConnachie2006, McConnachie2010} of the six GC candidates from \citet{beasley+2015} we have in our Gold sample. These clusters are compact, but have extended, asymmetrical shapes relative to a typical star. These cutouts are 10\arcsec\space on a side, and the blue circle indicates the 2\arcsec\space fiber diameter for APOGEE.}
\label{fig:beasley_pandas}
\end{figure*}

In Figure~\ref{fig:beasley_pandas}, we show the six GC candidates from \citet{beasley+2015} in our Gold sample. The PAndAS imagery of these objects exhibit extended or asymmetric profiles, evidence that these are not single stars and justifying their inclusion in our Gold sample. We also show images of the four new cluster candidates that we identify from their RVs that are distinct from the H\textsc{I} rotation curve in Section~\ref{sec:kinematics}. These clusters have compact and symmetric profiles, with the exception of AP01335532+3043580, indicating that despite their anomalous RVs, these are likely still MW foreground stars. AP01335532+3043580, however, looks strongly to be a cluster, and merits inclusion in further studies of the M33 cluster system.

\begin{figure*}
\centering
\includegraphics[width=0.8\textwidth]{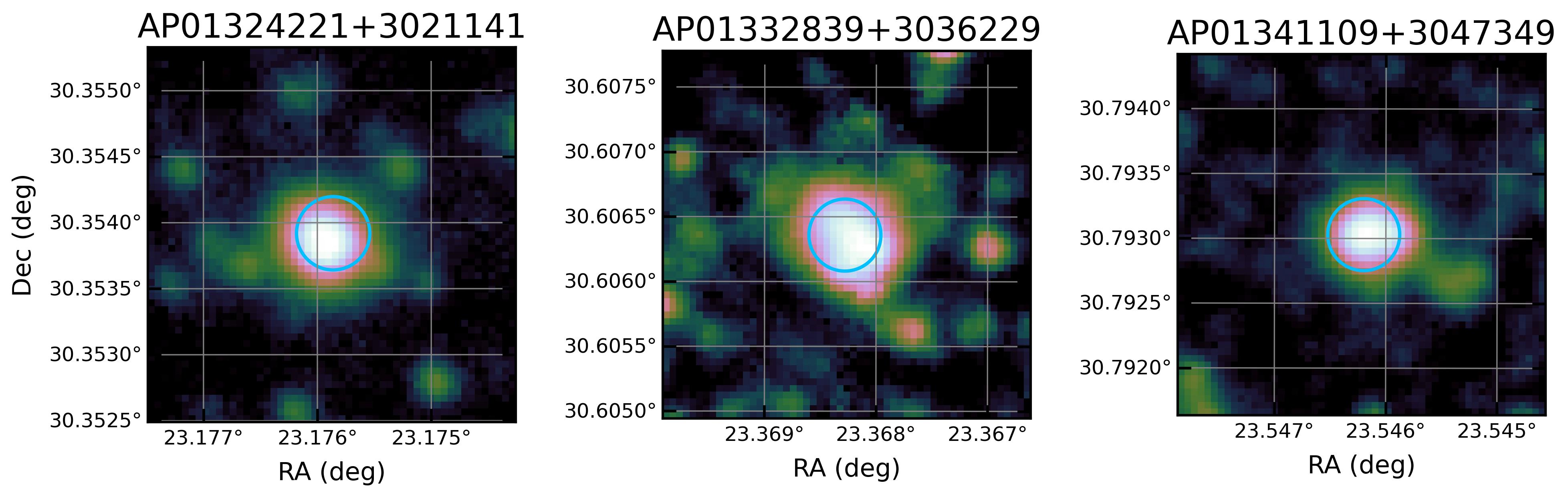}
\caption{Same as Figure~\ref{fig:beasley_pandas}, but for the three most metal rich clusters in our Gold sample. }
\label{fig:met_rich_pandas}
\end{figure*}

\begin{figure*}
\centering
\includegraphics[width=\textwidth]{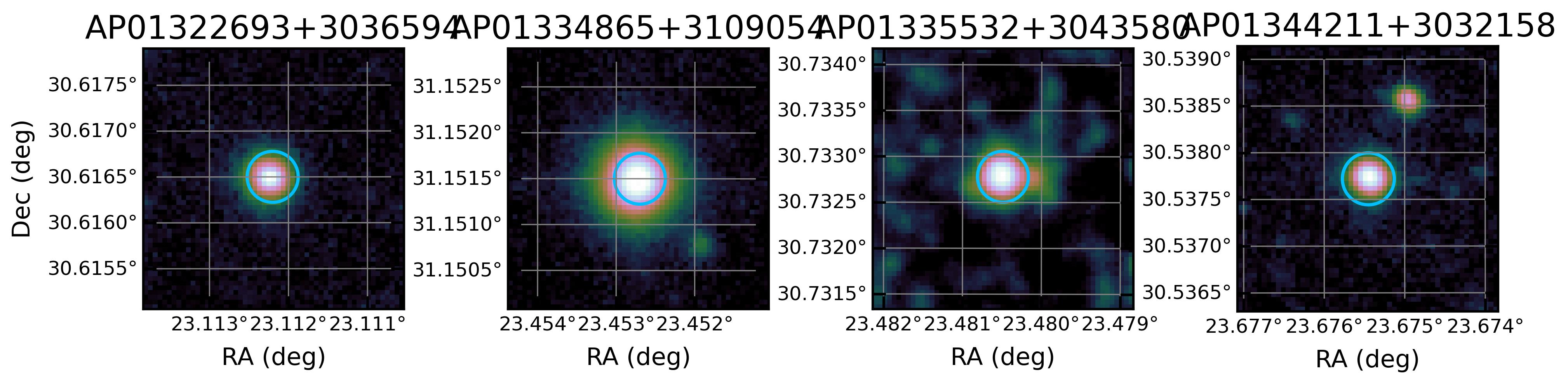}
\caption{Same as Figure~\ref{fig:beasley_pandas}, but for the four radial velocity cluster candidates mentioned in Section~\ref{sec:kinematics}.}
\label{fig:other_pandas}
\end{figure*}

\bibliography{biblio}{}
\bibliographystyle{aasjournalv7.1}



\end{document}